\documentclass[a4paper,11pt]{article}
\pdfoutput=1
\usepackage{jheppub}
\usepackage{amsmath}
\usepackage{amssymb}
\usepackage{dcolumn}	
\usepackage{bm}			
\usepackage{bbm} 		
\usepackage{multirow}
\usepackage{blkarray}
\usepackage{slashed}
\usepackage{bbding}
\usepackage{pifont}
\usepackage[usenames,dvipsnames]{xcolor}
\definecolor{hgreen}{rgb}{0,.3,0}
\definecolor{hred}{rgb}{.3,0,0}
\definecolor{hblue}{rgb}{0,0,.3}
\definecolor{LightGray}{gray}{0.95}
\makeatletter
\def\endfmffile{%
	\fmfcmd{\p@rcent\space the end.^^J%
		end.^^J%
		endinput;}%
	\if@fmfio
	\immediate\closeout\@outfmf
	\fi
	\ifnum\pdfshellescape>\z@
	\immediate\write18{mpost \thefmffile}%
	\fi}
\makeatother

\usepackage{pbox}
\usepackage{caption}
\usepackage{subcaption}
\usepackage{xcolor}
\usepackage[utf8]{inputenc}

\newcommand{\pid}{\hat{\pi}}
\newcommand{\fpid}{f_{\pid}}

\newcommand{\bmpsi}{\bm{m}_\psi}

\newcommand\ptwiddle[1]{\mathord{\mathop{#1}\limits^{\scriptscriptstyle(\bm{\sim})}}}

\begin{document}

\preprint{$\,$}

\title{Dark showers from $Z\,$-$\,$dark $Z'$ mixing}

\author[a]{Hsin-Chia Cheng,}
\affiliation[a]{Center for Quantum Mathematics and Physics (QMAP), Department of Physics, \\ University of California, Davis, CA, USA}
\author[b,c]{Xu-Hui Jiang,}
\author[d]{Lingfeng Li,}
\affiliation[b]{\mbox{Department of Physics, Hong Kong University of Science and Technology, Hong Kong}}
\affiliation[c]{Center for Future High Energy Physics, Institute of High Energy Physics, Chinese Academy\\ of Sciences and China Center of Advanced Science and Technology, Beijing, China}
\affiliation[d]{Department of Physics, Brown University, Providence, RI, USA}
\author[e,f]{Ennio Salvioni,}
\affiliation[e]{Dipartimento di Fisica e Astronomia, Universit\`a di Padova and\\ INFN, Sezione di Padova, Padua, Italy}
\affiliation[f]{Department of Physics and Astronomy, University of Sussex, Sussex House, Brighton, UK}
\date{\today}
\emailAdd{cheng@physics.ucdavis.edu, jiangxh@ihep.ac.cn, lingfeng\_li@brown.edu, E.Salvioni@sussex.ac.uk}
\abstract{
We discuss dark shower signals at the LHC from a dark QCD sector, containing GeV-scale dark pions. The portal with the Standard Model is given by the mixing of the $Z$ boson with a dark $Z^\prime$ coupled to the dark quarks. Both mass and kinetic mixings are included, but the mass mixing is the essential ingredient, as it is the one mediating visible decays of the long-lived dark pions. We focus especially on the possibility that the dark $Z'$ is {\it lighter} than the $Z$. Indirect constraints are dominated by electroweak precision tests, which we thoroughly discuss, showing that both $Z$-pole and low-energy observables are important. We then recast CMS and LHCb searches for displaced dimuon resonances to dark shower signals initiated by the production of on-shell $Z$ or $Z^\prime$, where the visible signature is left by a dark pion decaying to $\mu^+ \mu^-$. We demonstrate how dark shower topologies have already tested new parameter space in Run 2, reaching better sensitivity on a light dark $Z'$ compared to the flavor-changing decays of $B$ mesons, which can produce a single dark pion at a time, and the electroweak precision tests. 

}

\maketitle


\section{Introduction\label{sec:intro}}

A dark version of quantum chromodynamics (dark QCD), fully neutral under the Standard Model (SM) gauge symmetries and with a confinement scale of order $1$~GeV, appears in many new physics models. Notable examples include hidden valleys~\cite{Strassler:2006im}, strongly interacting massive particle (SIMP) dark matter~\cite{Hochberg:2014kqa}, as well as approaches to the hierarchy problem based on neutral naturalness~\cite{Chacko:2005pe} and cosmological relaxation~\cite{Graham:2015cka}. This scenario also has a strong appeal from a phenomenological perspective: provided some portal interactions connecting dark QCD and the SM exist, ``dark shower'' (DS) signals can be expected~\cite{Strassler:2006im,Schwaller:2015gea,Cohen:2015toa}. These pose fascinating new challenges to the LHC program and have received increasing attention lately, both from the theoretical side~\cite{Cohen:2017pzm,Pierce:2017taw,Beauchesne:2017yhh,Renner:2018fhh,Alimena:2019zri,Cheng:2019yai,Bernreuther:2019pfb,Cohen:2020afv,Knapen:2021eip,Linthorne:2021oiz,Cheng:2021kjg,Bernreuther:2022jlj,Albouy:2022cin,Cazzaniga:2022hxl,Beauchesne:2022phk,Born:2023vll,Carrasco:2023loy} and in actual experimental searches~\cite{Sirunyan:2018njd,Aaij:2020ikh,CMS:2021dzg,CMS:2021sch,ATLAS:2023swa,ATLAS:2023kao}.

If some dark quarks are lighter than the dark QCD confinement scale, the lightest dark hadrons are a set of dark pions, namely, the pseudo-Nambu-Goldstone bosons of chiral symmetry breaking generated by dark color dynamics. Unless the dark pions are stabilized by symmetries and act as dark matter candidates (e.g.~in SIMP models~\cite{Hochberg:2014kqa,Lee:2015gsa,Hochberg:2015vrg,Berlin:2018tvf,Katz:2020ywn} and related scenarios~\cite{Bernreuther:2019pfb,Bernreuther:2022jlj,Bernreuther:2023kcg}), their pattern of decays to SM particles, as mediated by the portal interactions, is crucial to identify the most promising search strategies. This is especially true for DS topologies, which are dominated by dark pions. 

The possibility that the $Z$ boson is the portal between the SM and dark QCD leads to a compelling and testable new physics scenario~\cite{Cheng:2021kjg} (built upon previous results~\cite{Cheng:2019yai}). In this $Z$ {\it portal} scenario some dark pions decay by mixing with the longitudinal mode of the $Z$, effectively behaving as composite axion-like particles (ALPs). For GeV-scale masses, their lifetimes fall between $1$~mm and $10$~m taking into account current constraints. Furthermore, DS events are produced by $Z$ decays at the LHC. The associated discovery potential is striking, thanks to the impressive statistics expected by the end of the high-luminosity phase (for example, about $2\times 10^{11}$ $Z$ bosons will be produced in each of ATLAS and CMS). Contrasting this with Higgs portal scenarios, the Higgs production cross section at the LHC is three orders of magnitude smaller than that of the $Z$ and current constraints require the lifetimes of dark pions decaying through Higgs mixing to be much longer.

The $Z$ portal can be described by the dimension-$6$ effective operator
\begin{equation}
\label{eq:Z_eff}
\mathcal{L}_\text{EFT} = \frac{1}{2} \left( \frac{g^2_{R,L}}{M_{\rm UV}^2} \right)_{ij} \big(i H^\dagger  D_\mu H + \mathrm{h.c.}  \big) \,\overline{\psi}_{R,L}^{\,i} \gamma^\mu  \psi^{\,j}_{R,L} \,,
\end{equation}
which in unitary gauge yields couplings between the dark quarks $\psi$ and the $Z$ boson (upon replacing the Higgs current in parentheses with $g_Z v^2 Z_\mu / 2$, where $v \approx 246$~GeV). Here $M_{\rm UV}$ and $g_{R,L}$ are the mass scale and couplings of the new physics that has been integrated out, assumed to be heavier than the weak scale. Such new physics also unavoidably contributes to the $T$ parameter~\cite{Peskin:1990zt,Peskin:1991sw} of electroweak precision tests (EWPT), associated to the operator $(i H^\dagger  D_\mu H + \mathrm{h.c.}  \big)^2$ (see also Ref.~\cite{Contino:2020tix} for a previous discussion).

\begin{figure}
\centering
\includegraphics[width=\textwidth]{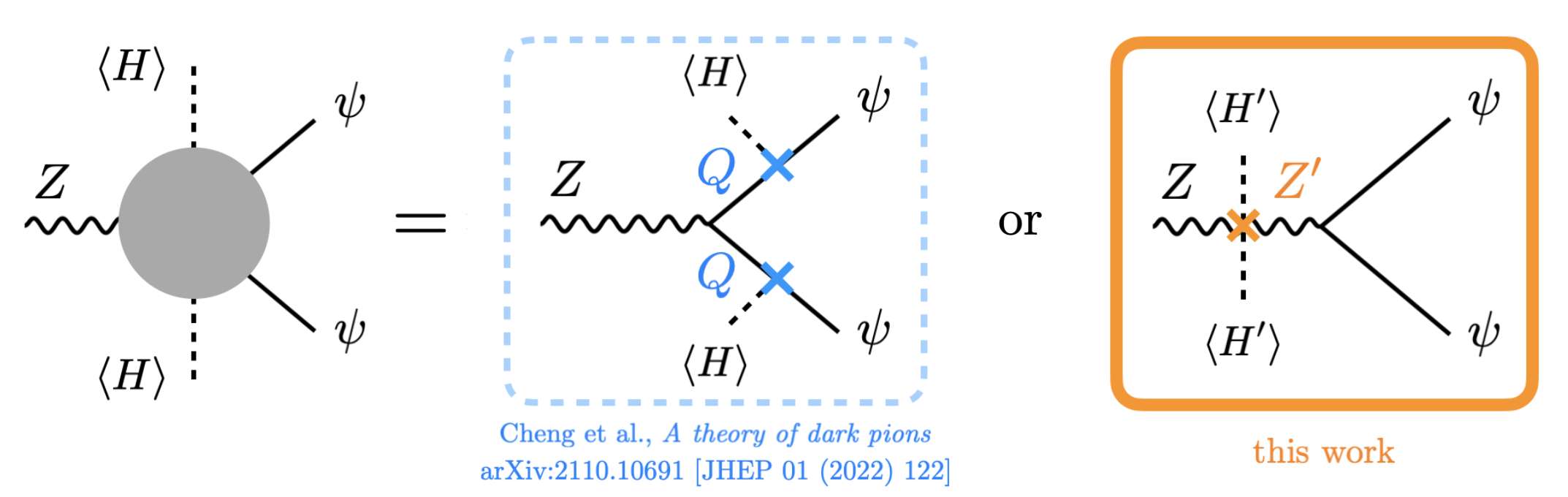}
\caption{The two classes of tree-level UV completions of the dimension-$6$ $Z$-portal operator in Eq.~\eqref{eq:Z_eff}. Completions with heavy fermions $Q$ were studied in Ref.~\cite{Cheng:2021kjg}. Here we consider completions with a dark $Z'$.\label{fig:mixing_diagrams}}
\end{figure}

Because the dark quarks are neutral under the SM gauge symmetries, mixings with ultraviolet (UV) states are required to generate the effective interaction Eq.~\eqref{eq:Z_eff} at tree level. As shown in Fig.~\ref{fig:mixing_diagrams}, two possible types of UV completions exist, where either the dark quarks or the $Z$ mix with new fields. Once we specify a UV completion, we can evaluate indirect constraints from EWPT and their interplay with direct searches at the LHC.

UV completions with dark quark mixing were studied in Ref.~\cite{Cheng:2021kjg}. The $\psi$ are coupled, via Yukawa couplings with the SM Higgs field, to other dark quarks $Q$ that also carry SM electroweak charges. Because of these charges, the $Q$ fermions must be heavier than the weak scale. In this case one has $g_{R,L}^2 / M_{\rm UV}^2 \simeq \ptwiddle{\bm{Y}} {}^{\,\dagger} \bm{M}^{-2} \ptwiddle{\bm{Y}}$, where $\bm{Y}$ and $\widetilde{\bm{Y}}$ are Yukawa matrices and $\bm{M}$ is the $Q$ mass matrix. The $T$ parameter is generated at one loop.

UV completions with $Z$ mixing are the subject of this work. A massive dark $Z'$ is introduced, associated to a spontaneously broken $U(1)^\prime$ under which the dark quarks are charged but the SM is neutral. Crucially, the $Z'$ has a {\it mass mixing} with the SM $Z$~\cite{Babu:1997st}; a kinetic mixing with hypercharge is also present in general, but it does not mediate the decay of dark pions via mixing with vector fields~\cite{Essig:2009nc}. The $Z\,$-$\,Z'$ mass mixing can be viewed as the leading low-energy effect of a second Higgs doublet $H'$ charged under $U(1)^\prime$~\cite{Davoudiasl:2012ag}, which acquires a vacuum expectation value (VEV). If the $Z'$ is heavier than the weak scale, the mass mixing yields $g_{R,L}^2 / M_{\rm UV}^2 \simeq (g_D \hat{g}_Z/ \hat{M}_{Z'}^2) ( \delta \hat{M}^2 / \hat{M}_Z^2 ) \bm{X}_{R,L}$, where $g_D$ is the $U(1)^\prime$ gauge coupling, $\bm{X}_{R,L}$ are the $\psi$ charges under this symmetry, and $\delta \hat{M}^2 / \hat{M}_Z^2$ is a dimensionless parameter quantifying the mass mixing. The $T$ parameter is generated at tree level.

However the $Z'$, being a SM singlet, does not necessarily need to be heavy;  its mass could be comparable to or even below the weak scale. In this paper we pay special attention to this possibility, and in particular to the mass range $10\;\mathrm{GeV}\lesssim M_{Z'} < M_Z$, where constraints from $B$ factories do not apply. We perform a thorough evaluation of the EWPT constraints on a dark $Z'$ with both mass and kinetic mixings, for arbitrary $M_{Z'}$, including $Z$-pole and low-energy data. The latter are shown to be important when $M_{Z'} \ll M_Z$. For a $Z'$ lighter than the $Z$, a mixing angle of $O(10^{-2})$ is allowed. These results do not depend on how the $Z'$ couples to the dark sector and are therefore of wider applicability.

A light $Z'$ can compete with or even dominate over the $Z$ as a source of DS events at the LHC. After an on-shell $Z$ or $Z'$ decays to dark quarks, dark parton shower and hadronization result in dark jets mainly composed of dark mesons. Assuming the masses of the dark vector mesons are larger than twice the dark pion masses, the vector mesons decay quickly and the dark jets end up being dominated by dark pions. The $CP$-odd dark pions decay by mixing with the longitudinal modes of $Z$ and $Z'$, provided a mass mixing is present, with macroscopic lifetimes. The $CP$-even dark pions decay by mixing with the Higgs and dark Higgs, but the resulting lifetimes are extremely long, so they can be viewed as effectively stable in laboratory experiments. 

In the type of DS events considered here, the production of a $Z$ or light $Z'$ is not automatically accompanied by other energetic objects. Therefore, the triggering strategies of LHC experiments need to exploit directly the dark pion ($\hat{\pi}$) decay products. Here we focus on dimuon displaced vertices from $\hat{\pi} \to \mu^+ \mu^-$, on which CMS and LHCb have exquisite sensitivity even at very small masses, $m_{\hat{\pi}}\sim O(1)$~GeV. CMS achieves this~\cite{CMS:2021sch} through scouting triggers, which enable higher rates by reducing the amount of information that is stored permanently. LHCb obtains competitive sensitivity~\cite{Aaij:2020ikh} by exploiting its low trigger thresholds and excellent vertex reconstruction~\cite{Pierce:2017taw}. Here we recast these searches to our parameter space. Another comparison of the capabilities of CMS and LHCb to discover DS in displaced dimuon events has appeared recently~\cite{Born:2023vll}, in the context of a different dark QCD model with a light dark photon. 

GeV-scale dark pions can also be singly-produced in flavor-changing neutral current (FCNC) decays of $B$ mesons. The displaced decays of dark pions to dimuons can be probed by the same CMS and LHCb searches~\cite{CMS:2021sch,Aaij:2020ikh}, which we recast to the FCNC signal as well. The FCNC sensitivity is compared to DS and to indirect bounds, which are dominated by EWPT. As we are going to show, DS topologies provide the best reach on a light $Z'$. In this work we report only current constraints, keeping to a minimum the discussion of the methods used to derive them. The methodology will be presented in a companion paper~\cite{Future:pheno}, where we also provide current bounds and future sensitivity projections, both in a model-independent approach and specializing to the two types of UV completions described above (heavy $Q$ fermions or dark $Z'$).

The remainder of this paper is organized as follows. In Sec.~\ref{sec:models} we introduce the model, including the dark QCD sector and the dark $Z'$ as the main mediator to the SM. In Sec.~\ref{sec:EWPO} we present our analysis of EWPT in the presence of a dark $Z'$ with both mass and kinetic mixings. Section~\ref{sec:other_bounds} collects other constraints on the setup, which turn out to be mostly subleading to EWPT. In Sec.~\ref{sec:dark_mesons} we develop the effective field theory (EFT) for dark mesons, including both dark pions and dark vector mesons $\hat{\rho}$. We describe decays of dark vector mesons to SM particles, which are important and phenomenologically promising if $m_{\hat{\rho}} < 2 m_{\hat{\pi}}$. We leave this region of parameter space for future study, focusing instead on $m_{\hat{\rho}} > 2 m_{\hat{\pi}}$ in the rest of the discussion. Finally, we identify a benchmark set of dark sector parameters, which we adopt in the phenomenological study contained in Sec.~\ref{sec:pheno}. There, we present our recasts of CMS and LHCb displaced dimuon searches to DS signals and FCNC $B$ meson decays. The results are compared to indirect constraints across the parameter space. Conclusions are drawn in Sec.~\ref{sec:Summary}. Five appendices (\ref{app:details},~\ref{app:potential},~\ref{app:EW_heavyZprime},~\ref{app:pheno_details}, and~\ref{app:DM}) provide additional details on both the model and the phenomenological analysis.

\section{Dark $Z'$ mediation to the dark sector}
\label{sec:models}

We assume that the dark quarks $\psi_j$ ($j = 1, \ldots, N$) are charged under a $U(1)^\prime$ gauge symmetry, in addition to the $SU(N_d)$ of dark QCD. The $U(1)^\prime$ is spontaneously broken by the VEV of a dark scalar $\Phi$, rendering the dark $Z'$ massive. The SM is completely neutral under $U(1)^\prime$, but interactions between the visible and dark sectors are mediated by the mixing of the $Z'$ and $\Phi$ with SM fields. We present here the salient features of the setup and refer to Appendix~\ref{app:details} for a detailed treatment.

In the spin-1 sector, the dark $Z'$ mixes with the SM gauge bosons. First, kinetic mixing with the hypercharge field is allowed. Second, mass mixing between the dark $Z'$ and $Z$~\cite{Babu:1997st} is possible if a second Higgs doublet $H^\prime$ is present, which carries $U(1)^\prime$ charge and obtains a nonzero VEV~\cite{Davoudiasl:2012ag,San:2022uud}.\footnote{The mass mixing can also arise from VEVs of scalars in larger representations of $SU(2)_L$. However, those are more constrained by custodial symmetry violation, so we only consider a Higgs doublet.} Such a VEV can be induced by a coupling between $H$, $H^\prime$, and $\Phi$ in the scalar potential, as discussed in Appendix~\ref{app:potential}. The $Z\,$-$\,Z^\prime$ {\it mass} mixing will turn out to be responsible for the decays of $CP$-odd dark pions.

In the spin-0 sector, assuming the $U(1)'$ is mostly broken by $\Phi$ and the second Higgs doublet $H'$ is relatively heavy, the physical scalars mostly contained in the latter decouple from the dark sector phenomenology below the weak scale, which is the focus of this work. Mixing of the $CP$-even scalars contained in $H$ and $\Phi$, on the other hand, will be responsible for the decays of $CP$-even dark pions. A priori, such mixing can be induced by $\Phi^\ast \Phi H^\dagger H$ or $H^\dagger H' \Phi^\ast + \text{h.c.}$ interactions. Upon integrating out the heavy $H'$ doublet, however, the latter operator ends up giving an additive contribution to the former at low energies (see Appendix~\ref{app:potential}). Therefore it is sufficient to focus on $\Phi^\ast \Phi H^\dagger H$.

Thus, the Lagrangian determining the mixing between the SM and the dark sector is
\begin{equation}
\mathcal {L}_{\rm mix} = -\frac{\sin \chi}{2} \hat{Z}'_{\mu\nu} \hat{B}^{\mu\nu} + \delta \hat{M}^2 \hat{Z}^{\prime \mu} \hat{Z}_\mu  - \kappa\, \Phi^\ast \Phi H^\dagger H \,.  \label{eq:Zhat2}
\end{equation}
We use hats to denote fields and couplings in this initial basis. The kinetic mixing between $U(1)_Y$ and $U(1)'$ is parametrized as $\sin \chi$, since the absolute value of this coefficient must be smaller than $1$ to ensure positivity of the kinetic energy. As already mentioned, the mass mixing $\delta \hat{M}^2$ (which we take to be positive without loss of generality) is implicitly assumed to originate from the VEV of the second Higgs doublet $H^\prime$; it represents the leading effect of $H'$ at low energies. This effective parametrization of a mass mixing between $Z$ and $Z'$ follows Ref.~\cite{Babu:1997st}. 

After performing a field redefinition to remove the kinetic mixing and a rotation to diagonalize the remaining mass mixing, we obtain the mass eigenstates
\begin{equation}\label{eq:mass_eigen}
\begin{pmatrix} A_\mu \\ Z_{\mu} \\ Z^\prime_{\mu} \end{pmatrix} = L^{-1}  \begin{pmatrix} \hat{A}_\mu \\ \hat{Z}_\mu \\ \hat{Z}'_\mu \end{pmatrix},
\end{equation}
where $A$ and $Z$ correspond to the physical photon and $Z$ boson, whereas $Z^\prime$ is the dark vector mass eigenstate. The mixing matrix is
\begin{align}
\label{eq:transform}
L &\;= \begin{pmatrix} 1 &\;\; -\hat{c}_W \sin \xi \tan \chi &\;\; - \hat{c}_W \cos \xi \tan \chi \\ 0 &\;\; \cos \xi + \hat{s}_W \sin \xi \tan \chi &\;\; -\sin \xi + \hat{s}_W \cos \xi \tan \chi \\ 0 &\;\; \sin \xi / \cos \chi &\;\;  \cos \xi / \cos \chi \end{pmatrix}\,,
\end{align}
with the mixing angle given by
\begin{equation}
\label{eq:mixing_angle}
\tan 2\xi = \frac{-2 \cos \chi ( \delta \hat{M}^2 + \hat{M}_Z^2 \hat{s}_W \sin \chi)}{\hat{M}_{Z'}^2 - \hat{M}_Z^2 \cos^2 \chi + \hat{M}_Z^2 \hat{s}_W^2 \sin^2 \chi + 2 \delta \hat{M}^2 \hat{s}_W \sin \chi}\,.
\end{equation}
Here $\hat{s}_W \equiv \sin \hat{\theta}_W$ and $\hat{c}_W \equiv \cos \hat{\theta}_W$, where $\hat{\theta}_W$ is the weak mixing angle in the limit where the dark sector is decoupled, whereas $\hat{M}^2_{Z}\,(\hat{M}^2_{Z^\prime})$ is the $Z\,$($Z^\prime$) squared mass in the initial field basis. The interactions of the neutral gauge boson eigenstates with the SM fermions and the dark quarks are
\begin{align}
- \left( \hat{e} J_{\rm EM}^\mu \;\;\; \hat{g}_Z \hat{J}_Z^\mu  \;\;\;  g_D {J}_D^\mu \right)  \begin{pmatrix} \hat{A}_\mu \\ \hat{Z}_{\mu} \\ \hat{Z}'_{\mu} \end{pmatrix}
= - \left( \hat{e} J_{\rm EM}^\mu \;\;\;  \hat{g}_Z \hat{J}_Z^\mu  \;\;\;  g_D {J}_D^\mu \right)  L \begin{pmatrix} A_\mu \\ Z_{\mu} \\ Z^\prime_{\mu} \end{pmatrix},
\end{align}
with $\hat{g}_Z= \hat{e} / (\hat{s}_W \hat{c}_W)$ and  
\begin{align}
J_{\rm EM}^\mu =  \bar{f} \gamma^\mu Q_f f \,,\qquad \hat{J}_Z^\mu =  \bar{f} \gamma^\mu ( T_{Lf}^3 - \hat{s}_W^2 Q_f ) f \,, \qquad {J}_D^\mu = \overline{\psi}_j \gamma^\mu x^j_{L(R)} P_{L(R)} \psi_j \,,
\end{align}
where sums over the (chiral) SM fermions $f$ and the dark flavor index $j$ are understood. The complete Lagrangian of the model and further details on the diagonalization to the physical states in the spin-1 sector are given in Appendix~\ref{app:details}.

Turning to the scalars, the VEVs $\langle H \rangle = (0 \;\; v/\sqrt{2})^T$ and $\langle \Phi \rangle = v_\Phi$ induce a mass mixing term between the radial modes $\hat{h}$ and $\hat{\phi}$, controlled by the quartic coupling $\kappa$. The mass eigenstates are obtained through the rotation
\begin{equation}
\begin{pmatrix} h \\ \phi \end{pmatrix} = \begin{pmatrix} \cos\theta_s & \sin\theta_s \\ - \sin\theta_s & \cos\theta_s \end{pmatrix} \begin{pmatrix} \hat{h} \\ \hat{\phi} \end{pmatrix} \,, \qquad
\tan 2 \theta_s = \frac{ 2 \sqrt{2} \kappa \hspace{0.15mm} v \hspace{0.15mm}  v_\Phi}{\hat{m}_h^2 - \hat{m}_\phi^2}\,,
\end{equation}
where $\hat{m}_{h,\,\phi}^2$ denote the diagonal elements of the scalar mass matrix in the initial field basis.
 
 \section{Electroweak precision constraints on a dark $Z'$ \label{sec:EWPO}}

The (mass and kinetic) mixings between the dark $Z'$ and the $Z$ affect electroweak observables. In this work we are especially interested in the possibility that the $Z^\prime$ is light -- in particular, lighter than the $Z$ -- so that an expansion in $M_{W}^2/M_{Z^\prime}^2 \ll 1$ is not justified. When such expansion is possible, a well-established and robust EFT framework is available~\cite{Barbieri:2004qk} to describe electroweak precision data in universal models like the one we consider here, where $\widehat{S}, \widehat{T}, W, Y$ are singled out as the leading beyond-the-SM (BSM) effects. On the contrary, for $M_{Z^\prime} \lesssim M_W$ one needs to proceed on a case-by-case basis. A light dark photon with pure kinetic mixing was thoroughly analyzed in previous studies~\cite{Hook:2010tw,Curtin:2014cca,Harigaya:2023uhg}. The case of a light dark $Z'$ with both mass and kinetic mixings, however, has received far less attention. A discussion has recently appeared in Ref.~\cite{Davoudiasl:2023cnc}; here we present an independent treatment, finding agreement where overlap is present. In this section we only provide the key aspects of our analysis, postponing more details to Appendix~\ref{app:EW_heavyZprime}, where we also explain the precise relation between the formalism described in this section and the general EFT parametrization of Ref.~\cite{Barbieri:2004qk}. Importantly, the results presented here are insensitive to the couplings of the $Z'$ to dark fermions. 

In our discussion, 
\begin{equation}\label{eq:inputs_SM}
\alpha\,,\;\; G_F\,,\;\; M_{Z}\,,
\end{equation}
are chosen as SM input parameters. The electromagnetic interaction is standard, so we have $\hat{e}=e$. On the other hand,
\begin{equation}
\label{eq:G_F}
\frac{G_F}{\sqrt{2}} = \frac{g^2}{8 M_W^2} = \frac{e^2}{8 \hat{s}_W^2 \hat{c}_W^2 \hat{M}_Z^2} = \frac{e^2}{ 8 s_W^2 c_W^2 M_{Z}^2}\,,
\end{equation}
where $s_W \equiv \sin \theta_W$, $c_W \equiv \cos \theta_W$ correspond to the weak mixing angle defined by
\begin{equation}
s_W^2 c_W^2 = \frac{\pi\hspace{0.2mm} \alpha (M_{Z})}{\sqrt{2}\hspace{0.4mm}G_F M_{Z}^2}\;,
\end{equation}
where $\alpha (M_{Z}) \approx (127.9)^{-1}$ is evaluated in the SM. 

\subsection{$Z$-pole observables}
The electroweak oblique parameters $S$, $T$, $U$~\cite{Peskin:1990zt,Peskin:1991sw} can be extracted from the Lagrangian describing the interaction of an on-shell $Z$ with the SM fermions,
\begin{equation} \label{eq:Z1ff}
\mathcal{L}_{Z f\bar{f}}= -\frac{ \bar{Z} e}{s_W c_W} \bar{f} \gamma^\mu ( T_{Lf}^3 - s_\ast^2 Q_f )  f Z_{\mu}\;,
\end{equation}
with
\begin{equation}\label{eq:s2star}
\bar{Z} =1 + \frac{\alpha T}{2}\,,\qquad  s_\ast^2 = s_W^2 + \frac{\alpha}{c_W^2- s_W^2} \Big(  \frac{S}{4} - s_W^2 c_W^2 T \Big)  ,
\end{equation}
as well as
\begin{equation}\label{eq:MW2}
\frac{M_W^2}{M_{Z}^2 } = c_W^2 + \frac{\alpha c_W^2}{c_W^2 - s_W^2} \Big( - \frac{S}{2}  + c_W^2  T   + \frac{ c_W^2 - s_W^2}{ 4 s_W^2} \,U  \Big)\,.
\end{equation}
In our approach, Eqs.~\eqref{eq:Z1ff},~\eqref{eq:s2star} and~\eqref{eq:MW2} should be viewed as {\it defining} $S, T, U$, which are therefore valid also for light new physics. These quantities also correspond to the BSM contributions to the $\epsilon_i$ parameters~\cite{Altarelli:1990zd}, with $\Delta \epsilon_1 = \alpha T,\; \Delta \epsilon_2 = - \alpha U / (4 s_W^2),\; \Delta \epsilon_3 = \alpha S / (4s_W^2)$. Their expressions in our model are given by Eqs.~\eqref{eq:Z1ff_model},~\eqref{eq:s_star} and~\eqref{eq:MW2model}, from which we obtain
\begin{align}\label{eq:PeskinTakeuchi}
S = \frac{4s_W^2}{\alpha} \left( \frac{c_W^2}{s_W} \xi t_\chi   -  c_W^2 \xi^2  \right)  \,, \quad T = \frac{1}{\alpha} \bigg( 2 s_W \xi  t_\chi + \xi^2  \bigg( \frac{M_{Z^\prime}^2}{M_{Z}^2} - 2 \bigg)  \bigg) \,, \quad
 U = \frac{4s_W^2}{\alpha} c_W^2 \xi^2\,,
\end{align}
where $t_\chi \equiv \tan \chi$. These results agree with Ref.~\cite{Holdom:1990xp}. We set constraints using the PDG fit to $S,T,U$~\cite{ParticleDataGroup:2022pth}.\footnote{Taking a conservative viewpoint, the fit does not include the recent $M_W$ measurement by CDF-II~\cite{CDF:2022hxs}.} We note that this fit is dominated by $Z$-pole observables (together with $M_W$), thus justifying the approach taken here. We also observe that the constraint disappears for $\xi = 0$, corresponding to $\sin \chi = -\, \delta \hat{M}^2 / (\hat{s}_W \hat{M}_Z^2)\,$. This flat direction is lifted by low-energy observables, as we now discuss.\footnote{Mixing between the $h$ and $\phi$ scalars also generates corrections to EW precision observables, at one-loop level. For instance, the contribution to $T$ can be found in Ref.~\cite{Profumo:2007wc}. For $m_\phi \gg m_h$ and omitting $O(M_V^2/m_h^2)$ corrections ($V = W, Z$) it yields the well-known~\cite{Barbieri:2007bh} logarithmic scaling $T = - 3 \sin^2 \theta_s \log \big(m_\phi^2 / m_h^2 \big) / (16\pi c_W^2)$, whereas in the opposite limit $m_\phi \ll m_h$ one obtains $T = 3 \sin^2 \theta_s m_h^2 \big[M_Z^2 \log (m_h^2 / M_Z^2) / (m_h^2 - M_Z^2) - \{ Z \to W \} \big] / (16\pi s_W^2 M_W^2)$. Throughout the parameter region considered here we find that this contribution to $|T|$ is $\lesssim 0.01$, much smaller than the corrections induced by the $Z\,$-$\,Z^\prime$ mixing presented above. Similar conclusions are reached for the $S$ and $U$ parameters. Possible small mixings with the $CP$-even component of $H^\prime$ are also neglected.}

\subsection{Low-energy observables} 
At energies $E \ll M_Z, M_{Z^\prime}\,$, one can integrate out both vectors to obtain an effective Lagrangian of current-current interactions,
\begin{equation}
\label{eq:Leff}
\mathcal{L}_{\rm eff} = -\frac{1}{2} \left( \hat{e} J_{\rm EM}^\mu \;\;\; \hat{g}_Z \hat{J}_Z^\mu \;\;\; g_D {J}_D^\mu \right)  L  \begin{pmatrix} 0 & & \\  & M_{Z}^{-2} &  \\  &  & M_{Z^\prime}^{-2} \end{pmatrix} L^T \begin{pmatrix} \hat{e} J_{{\rm EM} \,\mu} \\ \hat{g}_Z \hat{J}_{Z\mu} \\ g_D {J}_{D\mu} \end{pmatrix},
\end{equation} 
which we match to the four-fermion Lagrangian
\begin{equation}
\mathcal{L}_{\rm eff} = - \frac{4 G_F}{\sqrt{2}} \rho_\ast (0) \Big[ J_3^\mu J_{3\,\mu} - 2 s^2_\ast (0) J_3^\mu J_{\rm EM\,\mu} + O(J_{\rm EM}^2) \Big]\,,
\end{equation}
where $J_3^\mu =  \sum_f \bar{f} \gamma^\mu T_{Lf}^3 f\,$. We obtain
\begin{align} \label{eq:rhostar0}
\rho_\ast (0) =&\; 1 + \alpha T + \frac{M_{Z}^2}{M_{Z^\prime}^2} \big( \xi - s_W \tan\chi )^2 \,,\\
\frac{s_\ast^2 (0)}{s_W^2} =&\; \frac{s_\ast^2}{s_W^2} + \frac{M_{Z}^2}{M_{Z^\prime}^2} \Big( c_W^2 \tan^2 \chi - \frac{c_W^2}{s_W} \xi \tan \chi \Big)\,, \label{eq:s2star0}
\end{align}
where $ s_\ast^2 = s_\ast^2 (M_{Z}^2) $ is given by Eq.~\eqref{eq:s_star}. In Eqs.~\eqref{eq:rhostar0} and~\eqref{eq:s2star0}, the last term on each right-hand side corresponds to the contribution of the $Z^\prime$. These terms are enhanced when $M_{Z^\prime} < M_{Z}$.
 
The first low-energy observable we consider is atomic parity violation (APV). The relevant quantity is the weak charge of the Caesium atom, for which we find a theoretical prediction
\begin{equation} \label{eq:QW_prediction}
Q_W \approx Q_W^{\rm SM} \big( 1 + \delta \rho_\ast (0) + 2.91\, \delta s^2_\ast (0) \big)\,,
\end{equation}
where $\delta \rho_\ast (0) \equiv \rho_\ast (0) - 1$ and $\delta s^2_\ast (0) \equiv s^2_\ast (0) - s_W^2$. Our expression of $\delta Q_W$ as obtained from Eq.~\eqref{eq:QW} agrees with Eq.~(2.16) of Ref.~\cite{Altarelli:1991ci} for all the terms that were given there (our calculation retains some additional pieces).

The second observable is parity violation in $e^- e^- \to e^- e^-$ scattering at low energy, measured by the E158 experiment~\cite{SLACE158:2005uay}. It can be expressed in terms of the effective coupling
\begin{equation}\label{eq:ee_ee}
g_{AV}^{ee} = \rho_\ast (0) \Big( \frac{1}{2} - 2 s_\ast^2 (0) \Big)\quad \to \quad g_{AV}^{ee} \approx g_{AV}^{ee,\,\mathrm{SM}} \big( 1 + \delta \rho_\ast (0) - 87.0\,\delta s^2_\ast (0) \big)\,,
\end{equation}
whereas the third observable is the right-left asymmetry in $e^- p \to e^- p$ scattering as measured by the $Q_{\rm weak}$ experiment~\cite{Qweak:2018tjf}, which constrains the weak charge of the proton
\begin{equation}\label{eq:ep_ep}
g_{AV}^{ep} = \rho_\ast (0) \Big( -\frac{1}{2} + 2 s_\ast^2 (0) \Big)\quad \to \quad g_{AV}^{ep} \approx g_{AV}^{ep,\,\mathrm{SM}} \big( 1 + \delta \rho_\ast (0) - 87.0\,\delta s^2_\ast (0) \big)\,.
\end{equation}
As one can see, these two observables actually constrain the same combination of parameters.

\begin{figure}
\centering
\includegraphics[width=0.485\textwidth]{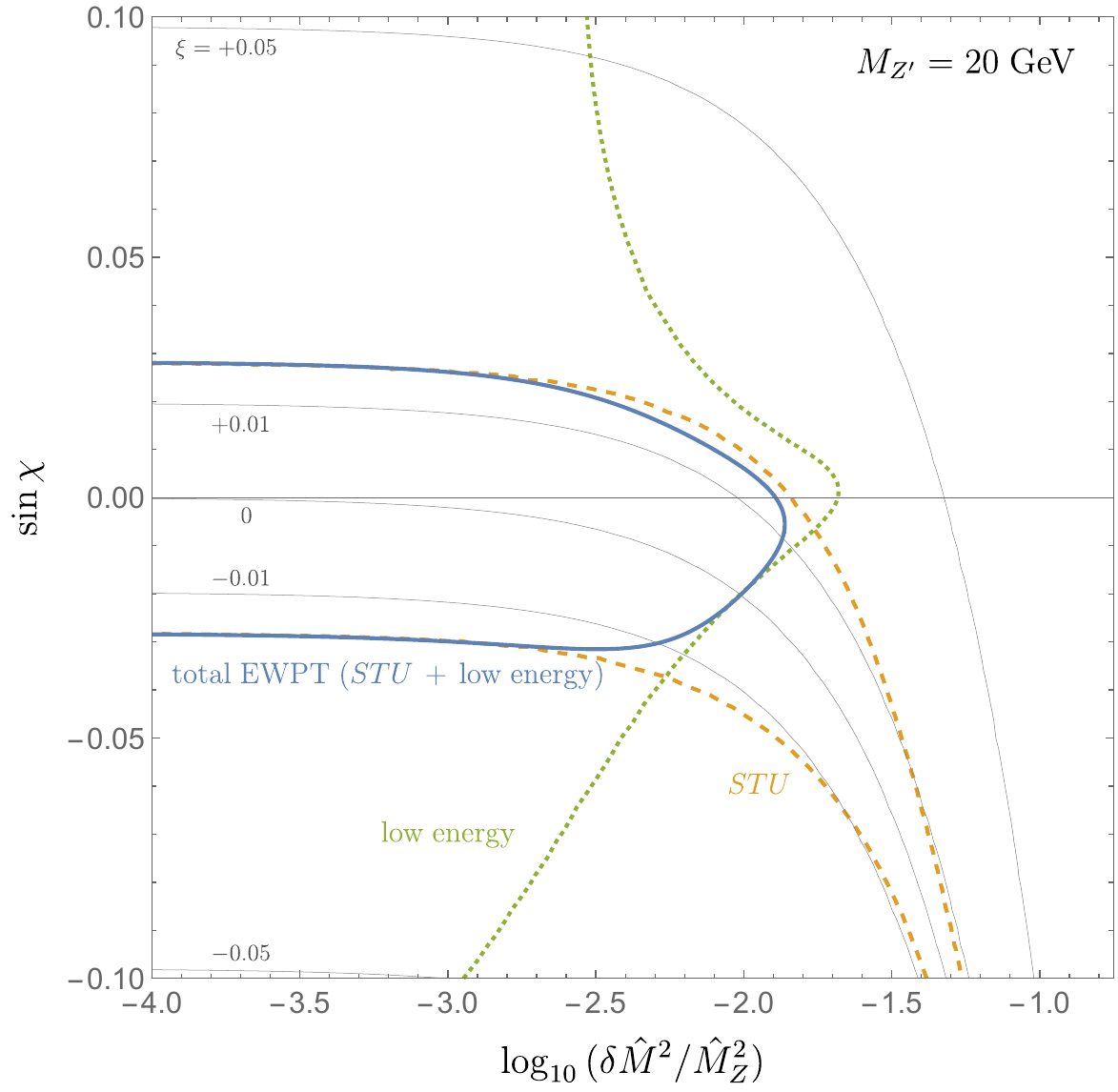}\hspace{2mm}
\includegraphics[width=0.485\textwidth]{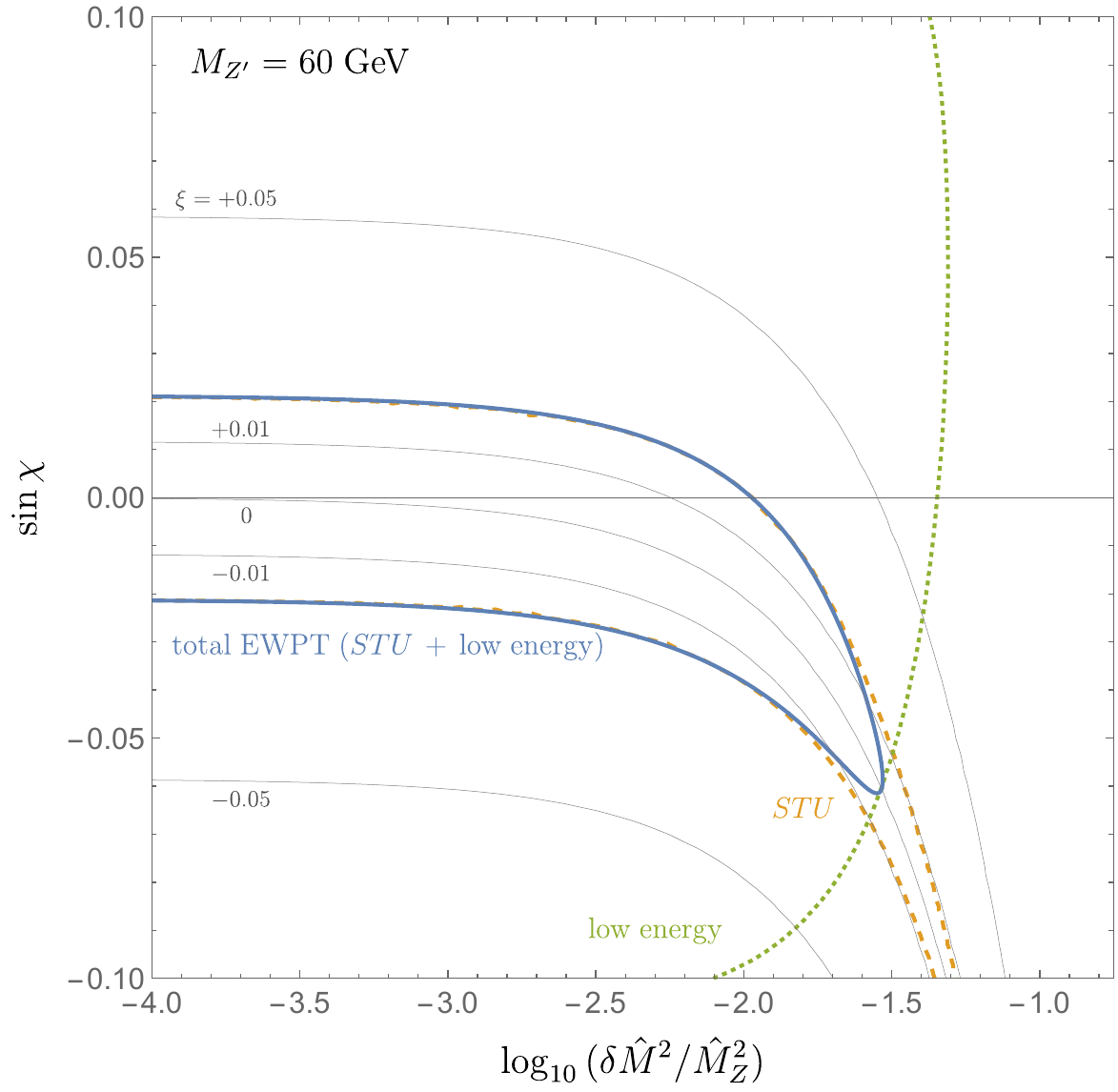}
\caption{$95\%$ CL constraints from electroweak precision data on the $Z\,$-$\,Z^\prime$ mass mixing parameter $\delta \hat{M}^2 / \hat{M}_Z^2$ and kinetic mixing parameter $\sin\chi$, for two representative benchmark values of the physical $Z^\prime$ mass: $M_{Z^\prime} = 20\;\mathrm{GeV}$ \emph{(left)} and $M_{Z^\prime} = 60\;\mathrm{GeV}$ \emph{(right)}. The dashed orange (dotted green) contour includes only the $STU$ fit (only the fit to low-energy data), whereas the solid blue contour combines the two. Iso-contours of the mixing angle $\xi$ are shown by thin gray curves.}
\label{fig:benchmarks}
\end{figure}

\subsection{Combination and bounds on the parameter space}
\begin{figure}
\centering
\includegraphics[width=0.495\textwidth]{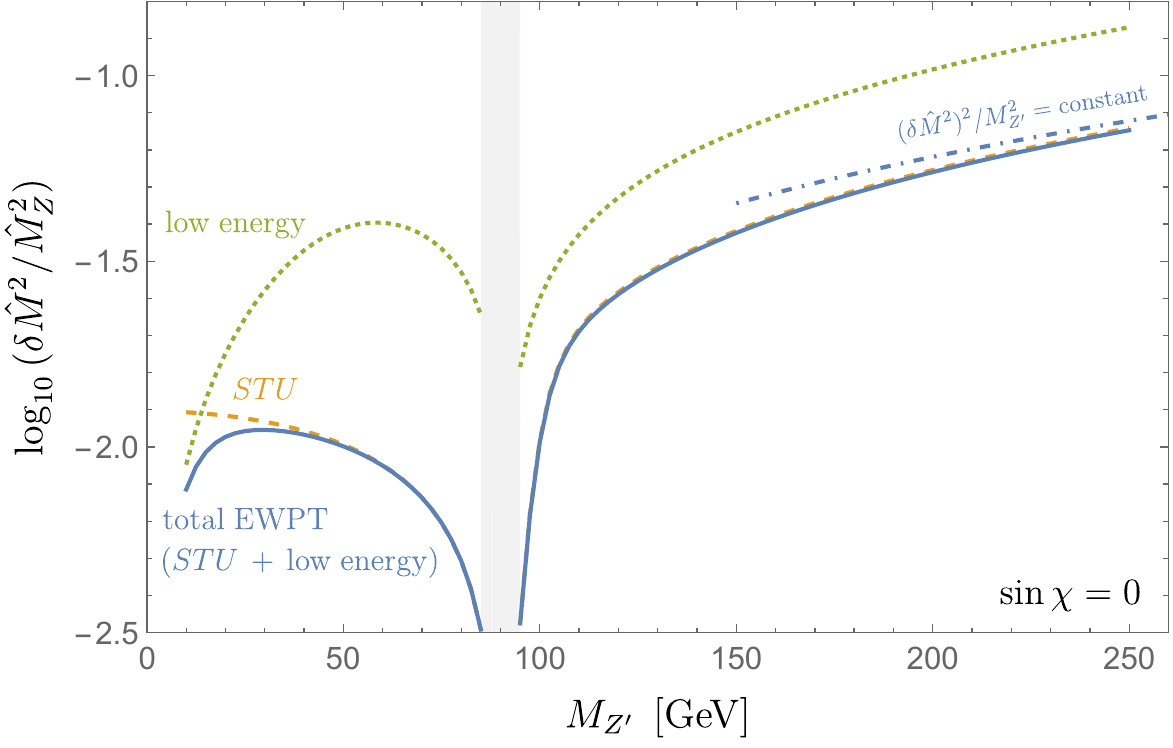}\hspace{0.25mm}
\includegraphics[width=0.495\textwidth]{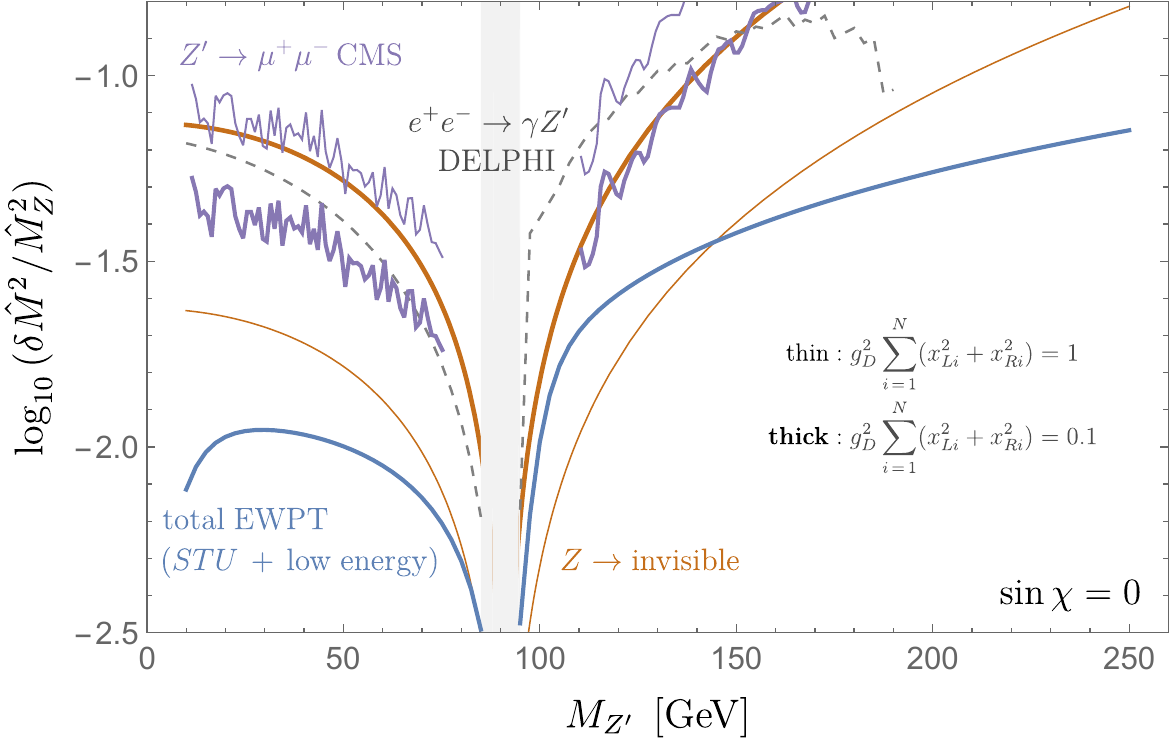}
\caption{{\it Left panel:} $95\%$ CL constraints on $\delta \hat{M}^2 / \hat{M}^2_{Z}$ from EW precision data assuming vanishing kinetic mixing, $\sin \chi = 0$. The dashed orange (dotted green) curve includes only the $STU$ fit (only the fit to low-energy data), whereas the solid blue curve combines the two. For large $M_{Z^\prime}$, the combined constraint approaches the EFT scaling $(\delta \hat{M}^2)^2/M_{Z^\prime}^2 = \mathrm{constant}$. {\it Right panel:} comparison of the EW precision constraints to other bounds. Purple lines indicate constraints from $Z^\prime \to \mu^+ \mu^-$ at CMS, for two representative values of the sum over dark quark $U(1)^\prime$ charges which controls the $Z'$ decay width to the dark sector. Brown lines show constraints from $Z\to \mathrm{invisible}$ at LEP, for the same two assumptions about dark quark charges, and the gray dashed line corresponds to $e^+ e^- \to \gamma (Z^\prime \to \mathrm{invisible})$ at DELPHI; these constraints from invisible final states apply only if the dark pion lifetimes are sufficiently long, $\tau_{\hat{\pi}} \gtrsim O(1)\,$m. We grayed out the region $M_{Z^\prime} \in [85, 95]$~GeV, where the new vector becomes near mass-degenerate with the $Z$.}
\label{fig:benchmark_sinchi_0}
\end{figure}

To set bounds on the parameter space we combine $Z$-pole and low-energy observables into a global $\chi^2$. Correlations between the low-energy observables (as well as between the $Z$-pole and low-energy ones) are neglected. We choose as input parameters, beyond those of the SM in Eq.~\eqref{eq:inputs_SM}, the quantities
\begin{equation}
 \sin \chi\,,\;\; M_{Z^\prime}\,,\;\; \delta \hat{M}^2 / \hat{M}^2_{Z}\, .
 \end{equation}
In Fig.~\ref{fig:benchmarks} we show the constraints on the $(\delta \hat{M}^2 / \hat{M}^2_{Z},\; \sin \chi)$ plane for two benchmark values of the $Z^\prime$ mass, $20$~GeV and 60~GeV, which we adopt throughout the discussion. The shape of the $Z$-pole exclusion clearly follows the $\xi = 0$ flat direction. It is cut off by the low-energy data, which are more constraining for lighter $Z'$ due to the presence of $M_{Z}^2 / M_{Z'}^2\,$-$\,$enhanced terms. In the left panel of Fig.~\ref{fig:benchmark_sinchi_0} we display the constraint on $\delta \hat{M}^2 / \hat{M}^2_{Z}$ as a function of the $Z^\prime$ mass, for vanishing kinetic mixing, again breaking down the $Z$-pole and low-energy components. The former dominates, except for $M_{Z'} \lesssim 30$~GeV where the latter becomes important. At large $Z'$ masses the effects on EWPT decouple as $(\delta \hat{M}^2)^2/M_{Z'}^2$ (see also Eq.~\eqref{eq:Barbieri_params}), which we find to be a good approximation already for $M_{Z'} \gtrsim 200$~GeV.

\section{Other constraints}\label{sec:other_bounds}
In this section we present additional constraints on the parameter space of the model, arising either from on-shell production of the $Z^\prime$, or from decays of the $Z$ and $h$ to the dark sector. We anticipate the results by pointing to the right panel of Fig.~\ref{fig:benchmark_sinchi_0}, which shows that these bounds are always subleading with respect to EWPT (except in a small window at $M_{Z'} \gtrsim M_Z$, where $Z\to \mathrm{invisible}$ can be competitive). 

Yet another source of constraints is, in principle, Deep Inelastic Scattering (DIS), which was exploited as a probe of kinetically mixed dark photons in Ref.~\cite{Kribs:2020vyk} and, like EWPT, is insensitive to the couplings of the new vector to dark sector states. While the recast to a $Z'$ with mass mixing is not immediate, we observe that for kinetic mixing DIS gives a weaker bound than EWPT in the region of masses we study in this paper, $m_{A'} \gtrsim 10$~GeV~\cite{Kribs:2020vyk}. Furthermore, for pure mass mixing the vector coupling of the $Z'$ to the electron is numerically suppressed (as it is for the $Z$), hence only the interference between the axial couplings of $Z$ and $Z^\prime$ is expected to be important, contrary to the dark photon model where photon$\,$-$A'$ interference dominates. Based on these considerations, we expect DIS to be less constraining than EWPT across our parameter space.

\subsection{On-shell $Z^\prime$ production at the LHC and LEP}
Owing to its sizeable couplings to dark quarks and small mixing with the SM, the dark $Z^\prime$ decays dominantly to dark hadrons. Two types of constraints on direct production of such a $Z^\prime$ need to be considered, for $M_{Z'} \gtrsim 10$~GeV: (i) searches for Drell-Yan production of dilepton resonances at the LHC~\cite{CMS:2019buh,LHCb:2019vmc}, whose exquisite sensitivity can compensate for the small $\mathrm{BR}(Z'\to \ell^+ \ell^-)$; and (ii) searches for $e^+ e^- \to \gamma Z^\prime$ production followed by $Z^\prime\to \mathrm{invisible}$ at LEP2~\cite{DELPHI:2003dlq}, which only apply if the dark pions (the dominant component of dark jets) are sufficiently long-lived, $\tau_{\hat{\pi}} \gtrsim O(1)\,\mathrm{m}$.

The physical $Z'$ interacts with the (Dirac) SM fermions as $(\hat{g}_Z / 2) \bar{f} \gamma^\mu (\bar{v}_f - \bar{a}_f \gamma_5) f Z_{\mu}^\prime\,$, where
\begin{align}
\bar{v}_f \,=&\,  (\sin \xi - \hat{s}_W \cos \xi \tan \chi ) T_{Lf}^3 -  \big[  (\sin \xi - \hat{s}_W \cos \xi \tan \chi ) \hat{s}_W^2 - \hat{s}_W \hat{c}_W^2 \cos \xi \tan \chi \big] 2 Q_f  \,, \nonumber\\
\bar{a}_f \,=&\, ( \sin \xi - \hat{s}_W \cos\xi \tan\chi ) T_{Lf}^3 \,, \label{eq:Z2_ff_couplings}
\end{align}
are the vector and axial-vector couplings.\footnote{For $\sin\chi = 0$ we have simply $\left\{\bar{v}_f, \bar{a}_f \right\} = \sin \xi \left\{v_f, a_f \right\}$, where $v_f = T_{Lf}^3 - 2 \hat{s}_W^2 Q_f$ and $a_f = T_{Lf}^3$ are the vector and axial-vector couplings of the SM $Z$.} The widths for the decays to dark quarks and SM fermions read
\begin{equation}
\Gamma (Z^\prime \to \overline{\psi}\psi, \bar{f} f) \,\simeq\; \frac{M_{Z^\prime}}{48\pi} \bigg\{ 2 N_d \frac{\cos^2 \xi}{\cos^2 \chi} g_D^2 \sum_{i\, =\, 1}^N (x_{L i}^2 + x_{R i}^2 )\,, \;  \hat{g}_Z^2 N_c (f) (\bar{v}_f^2 + \bar{a}_f^2 ) \bigg\}\,, \\
\end{equation}
where $g_D$ is the $U(1)^\prime$ gauge coupling and $x_{Li}\,$($x_{Ri}$) are the charges of the left-(right-)handed dark quarks. In all our numerical results we assume $N_d = 3$ dark colors. In the $Z'$ mass range we consider, decays to all SM fermions but the top quark are kinematically open. 

We derive constraints from the CMS search for dimuon resonances~\cite{CMS:2019buh} by adapting the results provided there for the standard dark photon model. Requiring the dimuon production cross sections to match in the two models (see Eq.~\eqref{eq:bound_Zprime_dileptons} in Appendix~\ref{app:onshell_Zprime}), we obtain bounds on the mass mixing parameter $\delta \hat{M}^2/\hat{M}_{Z}^2$ assuming a vanishing kinetic mixing. These are shown in the right panel of Fig.~\ref{fig:benchmark_sinchi_0}, for two representative choices of the sum over dark quark $U(1)'$ charges which determines the decay width of the $Z'$ to the dark sector, namely $g_D^2 \sum_{i\, =\, 1}^N (x_{L i}^2 + x_{R i}^2 ) = 1$~(thin purple line) and $0.1$~(thick purple line). To avoid clutter we do not show the constraints from the LHCb search for $A'\to \mu^+ \mu^-$~\cite{LHCb:2019vmc}, which has very similar sensitivity to CMS for $M_{Z'} \lesssim 70$~GeV.

Turning to searches for $Z^\prime \to \mathrm{invisible}$, for $M_{Z'} \gtrsim 10$~GeV the strongest sensitivity belongs to a DELPHI search for $e^+ e^- \to \gamma Z^\prime$ at LEP2 ($\sqrt{s}$ between 130 and 209 GeV)~\cite{DELPHI:2003dlq,Fox:2011fx}, from which bounds on a kinetically mixed dark photon with $\mathrm{BR}(A' \to \mathrm{invisible}) \approx 1$ were obtained in Refs.~\cite{Ilten:2018crw,Hochberg:2015vrg}. We reinterpret their results in our setup by requiring that the production cross sections match (see Eq.~\eqref{eq:DELPHI_limit}). The corresponding bound is shown by the dashed gray line in the right panel of Fig.~\ref{fig:benchmark_sinchi_0}. If we were to consider an even lighter $Z'$, far stronger constraints from $e^+ e^- \to \gamma Z^\prime$ production at BaBar~\cite{BaBar:2017tiz} would apply for $M_{Z'} \lesssim 8$~GeV.

\subsection{$Z$ and Higgs decays to the dark sector}
If the dark pions are sufficiently long-lived to escape detectors, decays of the $Z$ and Higgs bosons to invisible final states lead to additional constraints. The width for the $Z$ decay to dark quarks is
\begin{equation}\label{eq:Z_to_psipsibar}
\Gamma(Z \to \overline{\psi}\psi ) \simeq \frac{M_{Z}}{24\pi} N_d \frac{\sin^2 \xi}{\cos^2 \chi}\, g_D^2 \sum_{i \, =\, 1}^N (x_{L i}^2 + x_{R i}^2 ) \approx 0.36\;\mathrm{MeV} \, \Big( \frac{\xi}{0.01} \Big)^2 g_D^2 \sum_{i \, =\, 1}^N (x_{L i}^2 + x_{R i}^2 )\,.
\end{equation}
Decay to $Z^\prime \phi$ is also possible, if kinematically open, but this turns out to be typically suppressed by about one order of magnitude. The measurement of the $Z$ invisible width at LEP requires $\Delta \Gamma_Z^{\rm inv} < 2\;\mathrm{MeV}$ ($95\%$ CL)~\cite{ALEPH:2005ab}, which translates to the constraints shown by the brown lines in the right panel of Fig.~\ref{fig:benchmark_sinchi_0}.  

In addition, the $Z\,$-$\,Z^\prime$ mixing can induce the decay $h\to Z Z^\prime$ if $M_{Z^\prime}\lesssim 34$~GeV,
\begin{equation}\label{eq:h_Z_Zprime}
\Gamma(h \to Z Z^\prime) \simeq \frac{m_h^3 M_Z^2}{16\pi v^2 M_{Z^\prime}^2 } \big( \delta \hat{M}^2 / \hat{M}_Z^2 \big)^2  \left(\lambda_{h Z Z^\prime} + 12 \frac{M_Z^2 M_{Z^\prime}^2}{m_h^4} \right) \lambda_{h Z Z^\prime}^{1/2}\,,
\end{equation}
where we defined
\begin{equation}\label{eq:defs_Kallen}
\lambda_{ABC} \equiv \lambda (1, m_{B}^2/m_{A}^2, m_C^2/ m_{A}^2 )\,,\qquad \lambda (x, y, z) \equiv x^2 + y^2 + z^2 - 2 x y - 2 x z - 2 yz\,,
\end{equation}
leading to final states with both visible and invisible components. Contributions to Eq.~\eqref{eq:h_Z_Zprime} from scalar mixing are more suppressed and were neglected. We estimate
\begin{equation}
\frac{\mathrm{BR}(h \to Z\, (Z^\prime \to \mathrm{invisible}))}{\mathrm{BR}(h\to Z\, (Z^\ast \to \bar{\nu}\nu))_{\rm SM}} \approx 7\, \bigg( \frac{ \delta \hat{M}^2 / \hat{M}_Z^2 }{10^{-2}} \bigg)^2 \left( \frac{20\;\mathrm{GeV}}{M_{Z^\prime}} \right)^2\,.
\end{equation}
Thus the BSM rate can be larger than the SM one while remaining consistent with EWPT. However, even for the most promising decay mode, $Z\to \ell^+ \ell^-$ ($\ell = e, \mu$), the experimental prospects are challenging. The measurement of $h\to ZZ^{\ast} \to \ell^+ \ell^- + \slashed{E}_T$ is extremely difficult at the LHC, due to the very large $Z\,$+$\,$jets background for small values of the missing transverse energy. In fact, the analysis of this final state has only been performed targeting larger $h$ masses~\cite{CMS:2012qjb} or in the regime where $h$ is off shell~\cite{CMS:2022ley}. In conclusion, we believe that no significant bounds arise from this final state. Nevertheless, the $h\to Z Z^\prime$ decay could provide interesting sensitivity if the dark pions produced by the light $Z^\prime$ leave (displaced) visible signatures, since the $Z$ side of the event can ensure efficient triggering. We leave this for future investigation.

The mixing of scalars induces other decays of the Higgs to the dark sector, including $h \to \overline{\psi} \psi,\, \phi \phi,\, Z^\prime Z^\prime$. All of these are controlled by the scalar quartic coupling $\kappa$ in Eq.~\eqref{eq:Zhat2} and therefore more model dependent. A brief discussion can be found in Appendix~\ref{app:h_decays}.

\section{Effective theory for dark mesons}\label{sec:dark_mesons}

Having presented the existing constraints on the parameter space of the model, we now turn to describe the properties of the dark mesons, focusing in particular on their decays. For this purpose we adopt an EFT approach, where both the $Z$ and $Z^\prime$ are integrated out.

We assume that $N$ flavors of dark quarks, which transform as the fundamental representation of dark QCD, have masses below the confinement scale. Their $N\times N$ mass matrix ${\bmpsi}$ can in general contain also terms pairing dark quarks with different $U(1)^\prime$ charges, induced by Yukawa couplings $\bm{\zeta}^{1,2}$ to the dark scalar $\Phi$ (see Eq.~\eqref{eq:Zhat1} for the precise definitions); an explicit example is presented in Sec.~\ref{sec:benchmark}. Diagonalization is achieved via unitary transformations, $\psi_{L, R} = U_{L, R}\, \psi'_{L, R}\,$, where $\psi'_{L, R}$ are the mass eigenstates with diagonal mass matrix
\begin{equation}
{\bm{m}_{\psi'}} = U_L^\dag {\bmpsi} U_R\, .
\end{equation}
Then the $U(1)^\prime$ current
\begin{align}
J_D^\mu = \overline{\psi} \gamma^\mu \bm{X}_L P_L \psi + \overline{\psi} \gamma^\mu \bm{X}_R P_R \psi\,,
\end{align}
where $\bm{X}_L = \mathrm{diag}_i \left\{ x_{i L} \right\}$ and $\bm{X}_R = \mathrm{diag}_i \left\{ x_{i R} \right\}$,
is expressed in terms of the mass eigenstates as
\begin{align}\label{eq:dark_currents}
J_D^\mu = \overline{\psi'} \gamma^\mu \bm{X}'_V  \psi' + \overline{\psi'} \gamma^\mu \gamma_5 \bm{X}'_A \psi' 
\equiv J_{DV}^\mu + J_{DA}^\mu\,,
\end{align}
which defines the $N\times N$ matrices
\begin{equation}
\bm{X}'_{V,A} = \frac{1}{2} (\bm{X}'_R \pm \bm{X}'_L)\,, \qquad \bm{X}'_{L,R} = U_{L,R}^\dagger \bm{X}_{L,R} U_{L,R}\,. 
\end{equation}
Below the confinement scale, the current in Eq.~\eqref{eq:dark_currents} needs to be mapped to dark hadron multiplets. For $N> 1$ the lightest dark hadrons are expected to be a set of $N^2 - 1$ pseudo-Nambu-Goldstone bosons, the dark pions. Since they can be parametrically light and typically dominate the phenomenology, we begin by discussing their properties. Then we consider the vector mesons, whose masses are larger, in the ballpark of the dark confinement scale.

The low-energy interactions of the dark mesons with SM particles originate from the (dark $\times$ SM) piece of the four-fermion Lagrangian valid for $E\ll M_Z, M_{Z^\prime}$, Eq.~\eqref{eq:Leff}. When expressed in terms of vector and axial-vector currents, this yields
\begin{align} \label{eq:4f_SM_dark}
- \mathcal{L}_{\rm eff} & \supset  J_{DA}^\mu C_{AA}^{f}\bar{f}\gamma_\mu \gamma_5 f + J_{DA}^\mu C_{AV}^{f}\bar{f}\gamma_\mu  f + J_{DV}^\mu C_{VA}^{f}\bar{f}\gamma_\mu \gamma_5 f + J_{DV}^\mu C_{VV}^{f}\bar{f}\gamma_\mu  f \,,
\end{align}
where a sum over the SM fermions $f$ is understood, and
\begin{align}
C_{AA}^f =C_{VA}^f  &=  \frac{ g_D \hat{g}_Z a_f  \delta \hat{M}^2}{2 M_{Z}^2 M_{Z^\prime}^2  \cos ^2 \chi}\,, \\
C_{AV}^f =C_{VV}^f &= - \frac{ g_D}{2 M_{Z}^2 M_{Z^\prime}^2  \cos ^2 \chi} \left[ \hat{g}_Z  v_f \delta \hat{M}^2 + e\, 2 Q_f \hat{c}_W \hat{M}_Z^2 \sin \chi  \right]\,.
\end{align}
We now explore the consequences of these interactions for dark pions and dark vector mesons. For concreteness, we mostly focus on a theory with $N = 2$ dark flavors.

\subsection{Dark pions}
For $N=2$, the dark pions correspond to the fermion bilinears 
\begin{equation}
\hat{\pi}_a \sim i (\overline{\psi}_L^{\,\prime} \sigma_a \psi_R^\prime -  \overline{\psi}_R^{\,\prime} \sigma_a \psi_L^\prime) = \overline{\psi}^{\,\prime} \hspace{-0.5mm} i \sigma_a \gamma_5 \psi^\prime \,,
\label{eq:piondefinition}
\end{equation}
where $a \in \{1, 2, 3\}$ and $\sigma_a$ are the Pauli matrices. The $\hat{\pi}_{1,3}\; (\hat{\pi}_2)$ have $J^{PC} = 0^{-+}\; ( 0^{--})$ and are therefore $CP$-odd~(-even)~\cite{Cheng:2021kjg}. Although $CP$-violating phases can be present in general, leading to mixing between odd and even states, for simplicity in this work we focus on $CP$-conserving scenarios. The dark pions are created by the axial vector currents,
 \begin{equation} \label{eq:f_def}
\langle 0|j_{5a}^{\mu} (0) |\pid_b (p)\rangle = -\, i \delta_{ab} \fpid\,p^\mu\,, \qquad  j_{5a}^\mu= \overline{\psi}^{\,\prime}\gamma^\mu \gamma_5 \frac{\sigma_a}{2} \psi^\prime\,,
\end{equation}
with normalization of the decay constant $\fpid$ corresponding to $f_\pi \approx 93$ MeV in the SM. In Eq.~\eqref{eq:4f_SM_dark}, the only piece that mediates interactions of the dark pions with the SM is the one proportional to $C_{AA}^f$.~After replacing $J_{DA}^\mu$ with ${\rm Tr}(\sigma_b \bm{X}'_A) \fpid \partial^\mu \pid_b$ we obtain the effective Lagrangian that governs the decay of $CP$-odd dark pions through $Z\,$-$\,Z'$ mediation,
\begin{equation}\label{eq:piff}
\mathcal{L}_{\pid f\bar{f}}^{\,CP\,\text{-}\,\mathrm{odd}}  = -   \frac{\partial_\mu \pid_b}{f_a^{(b)}} \sum_f a_f \bar{f} \gamma^\mu \gamma_5 f\,,
\end{equation}
where the effective decay constant of the composite ALP $\pid_b$ is found by simple calculation,
\begin{equation}\label{eq:pion_decay_const}
f_a^{(b)} =   \frac{ 2 M_{Z}^2 M_{Z^\prime}^2  \cos ^2 \chi }{ {\rm Tr}(\sigma_b \bm{X}'_A) g_D  \hat{g}_Z \fpid  \delta \hat{M}^2 } \approx 1 \text{ PeV}\, \frac{ \cos ^2 \chi }{ {\rm Tr}(\sigma_b \bm{X}'_A) g_D } \left( \frac{ 1 \text{ GeV} }{ \fpid }\right) \bigg(\frac{ 10^{-2} }{\delta \hat{M}^2/ \hat{M}_{Z}^2 }\bigg)  \left( \frac{ M_{Z^\prime} }{60 \text{ GeV} } \right)^2 .
\end{equation}
Thus, dark pion decay is mediated by mass mixing between $Z$ and $Z'$, but {\it not} by kinetic mixing. This was expected~\cite{Essig:2009nc}: kinetic mixing only affects the transverse modes of gauge fields, and the exchange of a single transverse vector cannot mediate the decay of a pion.

Dark pion decays to SM particles mediated by the effective Lagrangian in Eq.~\eqref{eq:piff} have been thoroughly evaluated, as a function of $m_{\hat{\pi}}$ and  $f_a$, using a data-driven approach~\cite{Cheng:2021kjg}. Those results directly apply here. Very roughly, the decay to $\mu^+\mu^-$ dominates for $2m_\mu < m_{\hat{\pi}} \lesssim 800$~MeV, whereas for $800\;\mathrm{MeV} \lesssim m_{\hat{\pi}} \lesssim 3$~GeV the decay to $\pi^+ \pi^- \pi^0$ is largest, accompanied by a plethora of other modes with branching ratios at the $10\%$ level. For $m_{\hat{\pi}} \gtrsim 3.5$~GeV, decays to $c\bar{c}$ and $\tau^+ \tau^-$ have comparable branching ratios.

We pause briefly to offer two comments. First, if ${\rm Tr}(\sigma_a \bm{X}'_A)$ is small and therefore the direct decays of $CP$-odd dark pions to SM particles are suppressed, the $\hat{\pi}_a  \to \hat{\pi}_b\, f \bar{f}$ modes can be relevant, provided significant mass splittings between dark pions are present. These channels are mediated by the vector current, rather than the axial vector current; see Appendix~\ref{app:pia_pibff} for the calculation. Notice, however, that for $N = 2$ dark flavors the pion mass splittings are expected to be small, since they are absent at the leading order in the quark masses. Sizeable splittings are more likely for larger $N$.

Second, we wish to compare the expression of the dark pion decay constant $f_a$ obtained in the dark $Z'$ UV completion considered in this work, with its expression in the UV completion with heavy fermion mediators $Q$~\cite{Cheng:2021kjg}, which was already mentioned in Sec.~\ref{sec:intro}. Parametrically we find
\begin{equation}
\frac{(f_a)_{Q\, \rm fermions}}{(f_a)_{{\rm dark}\, Z'}} \sim \frac{\hat{g}_Z g_D}{Y^2} \frac{M^2}{M_{Z^\prime}^2} \frac{\delta \hat{M}^2}{\hat{M}_Z^2}\,,
\end{equation}
where we have assumed the traces over dimensionless matrices to be $O(1)$ numbers and, for simplicity, in the heavy fermion model we have set $\bm{\widetilde{Y}} = 0$. In the heavy fermion model, a Yukawa strength $Y \sim 1$ and vector-like $Q$ mass $M \sim 1\;\mathrm{TeV}$ are roughly at the edge of the region currently allowed by the $T$ parameter~\cite{Cheng:2021kjg}. Here, for a heavy $Z'$ with $M_{Z^\prime} = 1\;\mathrm{TeV}$ the electroweak constraints require $\delta \hat{M}^2 / \hat{M}_Z^2 < 0.30$ ($95\%$ CL, $\sin\chi = 0$). This cannot be compensated by increasing $g_D \gg 1$, which would lead to a rapid loss of perturbativity in the UV. Hence, for a heavy $Z'$ mediator EW precision constraints force $f_a$ to be mildly larger than it is allowed in a fermionic completion. However, the situation is reversed for light $Z'$: if $M_{Z^\prime} < M_{Z}$ the electroweak constraint $\delta \hat{M}^2 / \hat{M}_Z^2 \lesssim 10^{-2}$ (see Fig.~\ref{fig:benchmark_sinchi_0}) is more than compensated by the large $M^2/M_{Z^\prime}^2$ ratio. As a result, for a light $Z'$ the effective decay constant $(f_a)_{{\rm dark}\, Z'}$ can be more than one order of magnitude {\it smaller} than $(f_a)_{Q\, \rm fermions}$, while remaining consistent with EW precision constraints. Therefore, in the model studied here a larger portion of parameter space can be tested for the first time at the LHC (see the examination of the phenomenology in Sec.~\ref{sec:pheno}).

Resuming the discussion, besides $Z\,$-$\,Z^\prime$ mediation the decay of dark pions can proceed through the exchange of scalar fields. By integrating out $h$ and $\phi$ one obtains a low energy effective interaction between the dark quarks and SM fermions,
\begin{align}
 - \frac{y_f}{2\sqrt{2} } \frac{\sin\theta_s \cos \theta_s (m_h^2-m_\phi^2)}{\sqrt{2}\, m_h^2 m_\phi^2} \left[\overline{\psi}^{\,\prime} \Big( \bm{\zeta}^\prime + \bm{\zeta}^{\prime\, \dagger} + ( \bm{\zeta}^\prime - \bm{\zeta}^{\prime\, \dagger}) \gamma_5 \Big)  \psi'\right] \bar{f} f\,,
\end{align}
where $\bm{\zeta}^\prime = \, U_L^\dagger \bm{\zeta} U_R$ and $\bm{\zeta} \equiv \bm{\zeta}^1 + \bm{\zeta}^{2\dagger}$ is a combination of dark Yukawa matrices. From the relation (for $N=2$)
\begin{equation}\label{eq:B0hat}
\langle 0 | \overline{\psi}^{\,\prime} \frac{i \sigma_a}{2} \gamma_5 \psi' (0) | \pid_b (p)\rangle = - \delta_{ab} f_{\pid} \frac{ m^2_{\pid_a}}{ {\rm Tr} ( \bm{m}_{\psi'})} = - \delta_{ab} f_{\pid} \hat{B}_0\,,
\end{equation}
where $\hat{B}_0$ is a non-perturbative constant that we set for definiteness to $4\pi f_{\hat{\pi}}$, we derive an effective Lagrangian mediating the decay of $CP$-even dark pions,
\begin{equation}\label{eq:CP_even_decay}
\mathcal{L}_{\hat{\pi}f\bar{f}}^{\,CP\,\text{-}\,\mathrm{even}} = - s_{\theta}^{(b)} \frac{m_f}{v} \pid_b \sum_f \bar{f} f\,.
\end{equation}
We thus obtain the effective mixing angle
\begin{align}\label{eq:pion_mix_angle}
s_{\theta}^{(b)} =\,& \frac{2\pi f_{\hat{\pi}}^2 \kappa v_\Phi v }{m_h^2 m_\phi^2}\, {\rm Tr} [ i \sigma_b (\bm{\zeta}^\prime - \bm{\zeta}^{\prime\, \dagger} ) ]  \\ 
\approx\,& 10^{-8} \frac{1}{g_D | x_\Phi |} \bigg( \frac{f_{\hat{\pi}}}{1\;\mathrm{GeV}} \bigg)^2 \bigg(\frac{\kappa}{0.1}\bigg) \bigg( \frac{M_{Z^\prime}}{60\;\mathrm{GeV}} \bigg) \bigg( \frac{200\;\mathrm{GeV}}{m_\phi} \bigg)^2 \bigg( \frac{{\rm Tr} [ i \sigma_b (\bm{\zeta}^\prime - \bm{\zeta}^{\prime\, \dagger} ) ]}{10^{-3}} \bigg) \,, \nonumber
\end{align}
where $M_{Z^\prime} \simeq \hat{M}_{Z'} \simeq \sqrt{2}\, g_D | x_\Phi | v_\Phi\,$ has been used in the second line. Recalling that  $m^2_{\hat{\pi}} = 4 \pi f_{\hat{\pi}} \mathrm{Tr} (\bm{m}_{\psi^\prime} )$, the chosen normalization for the trace corresponds to having a dark pion mass $m_{\hat{\pi}} \sim 1$~GeV for $f_{\hat{\pi}} \sim 1$~GeV and $v_{\Phi} \sim 100$~GeV, when the $O(\bm{\zeta} v_\Phi)$ terms originating from the dark Yukawas contribute at order one to the dark quark mass matrix. If a significant hierarchy is present between $\bm{\zeta} v_\Phi$ and the ``bare'' dark quark masses $\bm{m}$, the effective mixing angle $s_\theta^{(b)}$ is even more suppressed.

Decays to SM particles mediated by the Lagrangian in Eq.~\eqref{eq:CP_even_decay} have been extensively studied~\cite{Winkler:2018qyg}, and we simply make use of previous results~\cite{Winkler:2018qyg,Cheng:2021kjg}. For $m_{\hat{\pi}} \lesssim 2\;\mathrm{GeV}$, decays to $\pi\pi$ or $K\overline{K}$ dominate.

 \subsection{Dark vector mesons}

Higher in the dark hadron spectrum we expect vector mesons ($\hat{\rho}$ and $\hat{\omega}$) to appear. If their masses are larger than $2m_{\hat{\pi}}$, after they are produced the vector mesons decay quickly to dark pions (assuming dark isospin breaking is not too small, so that the decay of $\hat{\omega}$ to $\hat{\pi}\hat{\pi}$ is not too suppressed). This is the scenario we focus on in this paper, where the dark pions dominate the DS phenomenology. Nonetheless, we offer here a first discussion of the complementary scenario, where the decays to dark pions are kinematically closed and the vector mesons decay to SM particles via mixing with the $Z$ or $Z'$. This scenario is phenomenologically interesting in its own right. In particular, if $\mathrm{Tr}(\sigma_b \bm{X}'_A) \ll \mathrm{Tr}(\sigma_b \bm{X}'_V)$ the dark pions can be effectively stable at colliders, while most visible signals would originate from the vector mesons.  

\begin{figure}[t]
\centering
\includegraphics[width=0.495\textwidth]{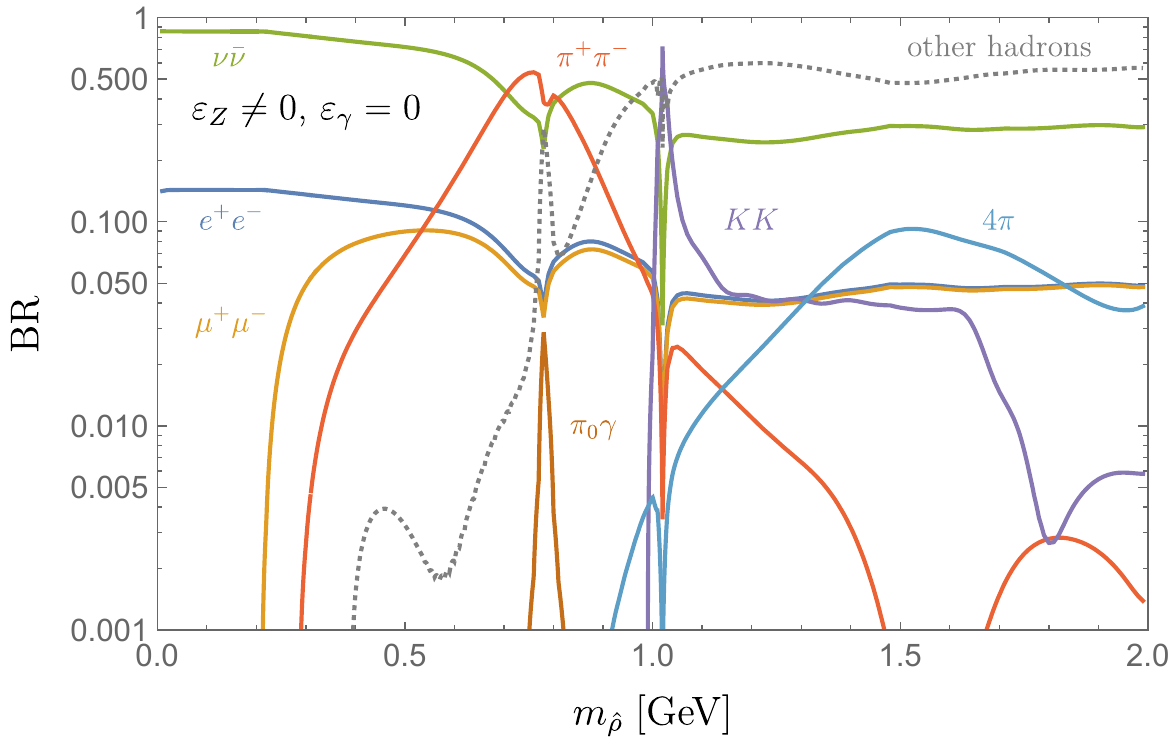}\hspace{0.25mm}
\includegraphics[width=0.4925\textwidth]{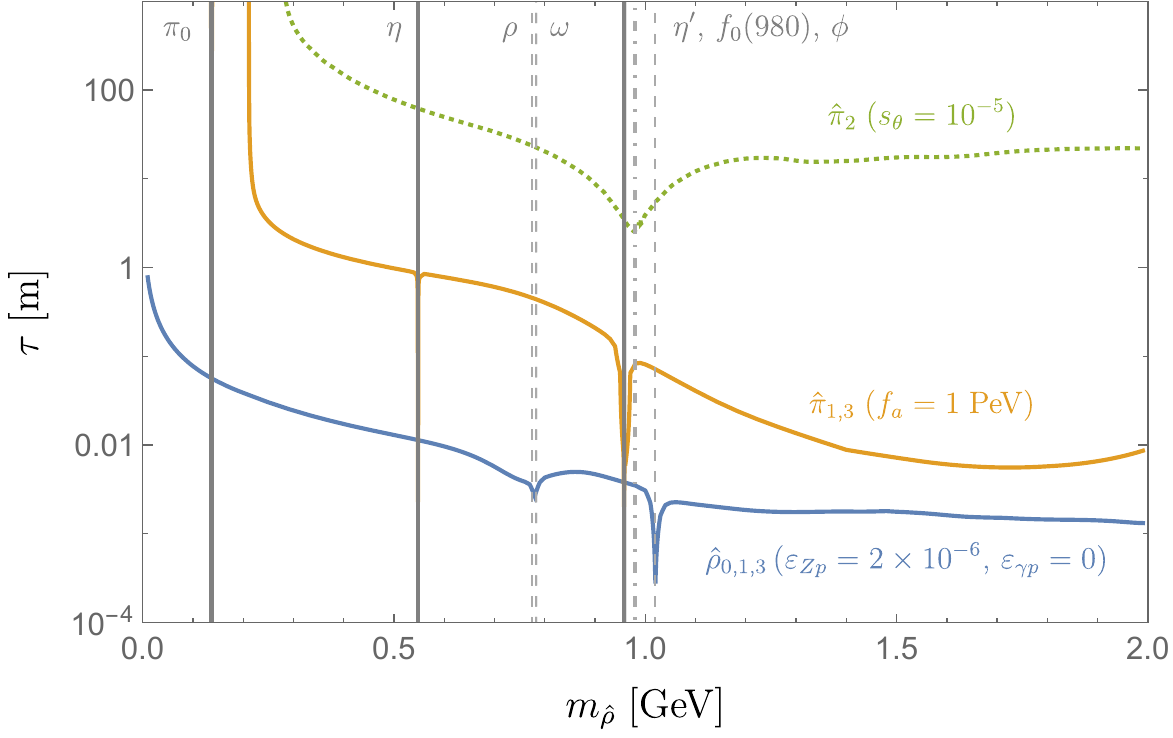}
\caption{{\it Left panel:} Branching ratios for a $CP$-even $\hat{\rho}$ coupled only to the $Z$ current ($\varepsilon_{\gamma} = 0$), obtained from the DarkCast package~\cite{Ilten:2018crw,Baruch:2022esd}. We show separately a set of exclusive hadronic decays mediated by the vector current, whereas all other hadronic modes are merged and indicated by the dotted gray curve. {\it Right panel:} Proper lifetime of a $CP$-even $\hat{\rho}_p$ coupled to the $Z$ current, for a typical value of $\varepsilon_{Zp}$ (solid blue). For comparison we consider the $CP$-odd dark pions with a typical value of $f_a$ (solid orange). The $CP$-even dark pion is also displayed (dotted green), choosing $s_\theta = 10^{-5}$ to fit within the plot range, though the typical expected value is far smaller, see Eq.~\eqref{eq:pion_mix_angle}. In both panels, only decays to SM particles are considered.}\label{fig:rho_BRs} 
\end{figure}

The vector mesons are created by vector currents. At hadron level, these are mapped to
\begin{equation}
j_{p}^\mu = \overline{\psi}^{\,\prime} \gamma^\mu \frac{\sigma_p}{2} \psi^\prime\;\; \to \;\; -\, \frac{m_{\hat{\rho}}^2}{g_{\hat{\rho}}}\,\hat{\rho}_{p}^{\,\mu}\; , \qquad \hat{\rho}_{p}^{\,\mu} = (\hat{\omega}^\mu, \hat{\rho}_{a}^{\,\mu})\,,
\end{equation}
where $p \in \{0, 1, 2 , 3\}$ and $\sigma_0 \equiv \mathbf{1}_2$. Here $g_{\hat{\rho}}$ is the vector meson coupling, which in the SM takes the value \mbox{$g_\rho \simeq m_\rho / (\sqrt{2} f_\pi) \approx 6\,$}. The $\hat{\rho}_{\hspace{0.3mm}0,1,3}$ have $J^{PC} = 1^{--}$ whereas $\hat{\rho}_2$ has $1^{-+}$, as can be derived using $\overline{\psi}^{\,\prime}_2 \gamma^\mu P_{L,R} \psi_1^\prime \stackrel{C} {\to} -\, \overline{\psi}^{\,\prime}_1 \gamma^\mu P_{R,L} \psi^\prime_2\,$. Replacing $J_{DV}^\mu$ in Eq.~\eqref{eq:4f_SM_dark} by $- {\rm Tr} (\sigma_p \bm{X}'_V) m_{\hat{\rho}}^2\, \hat{\rho}_{p}^{\,\mu} / g_{\hat{\rho}}\,$ leads to the following interactions with SM fermions,
\begin{equation}
\mathcal{L}_{\hat{\rho} f \bar{f}}  =  {\rm Tr} (\sigma_p \bm{X}'_V) \frac{ m_{\hat{\rho}}^2} {g_{\hat{\rho}}} \hat{\rho}_{p}^{\,\mu}  \bar{f} \gamma^\mu \big( C_{VV}^f  + C_{VA}^f \gamma_5 \big) f  =  -\, \hat{\rho}^{\,\mu}_p \left( \varepsilon_{\gamma p} e J^\mu_{{\rm EM}} + \varepsilon_{Z p}\, \hat{g}_Z \hat{J}_Z^\mu \right),
\end{equation}
where in the second expression a sum over $f$ is understood, and we have defined 
\begin{align}
\varepsilon_{\gamma p}&=   g_D {\rm Tr} ( \sigma_p \bm{X}'_V)  \frac{ \hat{M}_Z^2 \hat{c}_W \sin \chi }{M_{Z}^2 M_{Z^\prime}^2 \cos^2 \chi }    \frac{m_{\hat{\rho}}^2}{g_{\hat{\rho}}}  \\
&\approx  1.7 \times 10^{-6}\; \frac{g_D {\rm Tr} ( \sigma_p \bm{X}'_V) }{\cos^2 \chi} \left( \frac{ 60 \text{ GeV}}{M_{Z^\prime}}\right)^2  \left(  \frac{m_{\hat{\rho}}^2}{ 2 \text{ GeV}} \right)^2 \left( \frac{6}{g_{\hat{\rho}}} \right) \left( \frac{ \sin \chi}{10^{-2}}\right) ,  \nonumber\\
\varepsilon_{Z p} &=  g_D {\rm Tr} ( \sigma_p \bm{X}'_V)  \frac{ \delta \hat{M}^2}{ M_{Z}^2 M_{Z^\prime}^2 \cos^2 \chi }   \frac{m_{\hat{\rho}}^2}{g_{\hat{\rho}}}  \\
&\approx  1.9 \times 10^{-6} \; \frac{g_D {\rm Tr} ( \sigma_p \bm{X}'_V) }{\cos^2 \chi} \left( \frac{ 60 \text{ GeV}}{M_{Z^\prime}}\right)^2  \left(  \frac{m_{\hat{\rho}}^2}{ 2 \text{ GeV}} \right)^2 \left( \frac{6}{g_{\hat{\rho}}} \right) \left(\frac{\delta \hat{M}^2 /\hat{M}_{Z}^2}{10^{-2}}\right) . \nonumber
\end{align}
Thus, the dark vector mesons decay back to SM charged leptons and neutrinos with
\begin{equation}
\Gamma(\hat{\rho}_{p}\to f\bar{f}) = \frac{m_{\hat{\rho}_{p}} }{12\pi}  \lambda_{\hat{\rho}_p f \bar{f}}^{3/2} \Bigg[ \hspace{-1mm} \left(\varepsilon_{\gamma p} e Q_f  + \varepsilon_{Z p} \frac{\hat{g}_Z}{2} v_f \right)^2 \hspace{-1mm} \bigg( 1  + \frac{6m_f^2}{m_{\hat{\rho}_{p}}^2 \lambda_{\hat{\rho}_p f \bar{f}}} \bigg) + \left( \varepsilon_{Z p} \frac{\hat{g}_Z}{2} a_f \right)^2 \hspace{-0.5mm}\Bigg]. 
\end{equation} 
For decays to SM hadrons, a general discussion of new light spin-1 particles with arbitrary vector and axial-vector couplings to SM quarks was presented in Refs.~\cite{Ilten:2018crw,Baruch:2022esd} (see also Ref.~\cite{Foguel:2022ppx}). In the left panel of Fig.~\ref{fig:rho_BRs} we show the branching ratios to SM final states for $\hat{\rho}_p$ interacting only with the $Z$ current ($\varepsilon_{\gamma p} = 0$), which is characteristic of our scenario. The branching ratios were obtained using the DarkCast code.\footnote{In Eq.~(2.8) of~\cite{Baruch:2022esd} and Eq.~(2.13) of~\cite{Ilten:2018crw}, $\mathcal{C}_\nu$ should equal $1$ instead of $1/2$. This has been corrected in presenting our results, as well as in the latest release of DarkCast (v2.1).} We observe that a large branching ratio for dark vector decay to $\nu\bar{\nu}$ is a distinctive feature. In the right panel of Fig.~\ref{fig:rho_BRs}, we compare the lifetime of the dark vector mesons with those of the dark pions. It can be seen that the lifetime of the $CP$-even vector mesons $\hat{\rho}_{\hspace{0.3mm}0,1,3}$ is one to two orders of magnitude smaller than the shortest $\hat{\pi}$ lifetime, for typical values of the parameters. By contrast, if $CP$-violating phases are absent $\hat{\rho}_2$ does not decay through the above mechanism, because it cannot mix with a $1^{--}$ or $1^{++}$ state.\footnote{It is easy to verify this by direct calculation: exploiting the hermiticity of $\bm{X}^\prime_V$ one finds $\mathrm{Tr}(\sigma_2 \bm{X}^\prime_V) \propto (X^\prime_{V})_{12} - (X^\prime_V)_{12}^\ast = 0\,$.} For reference, we note that the lightest SM meson with quantum numbers matching those of $\hat{\rho}_2$ is the $\pi_1(1400)$. 

Another possible decay mechanism when $m_{\hat{\rho}} < 2 m_{\hat{\pi}}$ is $\hat{\rho} \to \hat{\pi} (\hat{\pi}^\ast \to \mathrm{SM})$. For the isospin triplet vectors, the $\hat{\rho}_a \hat{\pi} \hat{\pi}$ coupling is given by
\begin{equation}
i g_{\hat{\rho}} \,\mathrm{Tr} \big( \hat{\rho}_a^\mu \sigma_a  [ \hat{\Pi} , \partial^\mu \hat{\Pi} ] \big) / 4 \; = \; -\, g_{\hat{\rho}}\, \epsilon^{abc} \hat{\rho}_a^\mu \hat{\pi}_b \partial_\mu \hat{\pi}_c \qquad \big(\hat{\Pi} = \hat{\pi}^b \sigma_b \big)\,.
\end{equation}
This may be an important (or even dominant) decay mode for $\hat{\rho}_2$, whose direct decay to SM particles is likely very suppressed. For $\hat{\rho}_{1,3}$, it competes with the direct $\hat{\rho}_{1,3}\to \mathrm{SM}$ channels. For the isospin singlet vector, the $\hat{\omega}\hat{\pi}\hat{\pi}$ coupling is proportional to isospin breaking. The study of the dark vector meson phenomenology in the $m_{\hat{\rho}} < 2 m_{\hat{\pi}}$ scenario, for which the above discussion of decays to SM final states is central, is left for future work.

\subsection{A benchmark dark sector}\label{sec:benchmark}
The traces appearing in Eqs.~\eqref{eq:pion_decay_const} and~\eqref{eq:pion_mix_angle} encapsulate the imprint on the dark pions of the structure and symmetries of the underlying quark theory. For concreteness, we focus on a simple and natural benchmark model with $N = 2$ dark flavors. The dark quarks $\psi_{1,2}$ have vector-like charges under $U(1)'$, called $x_{1,2}$, and the scalar $\Phi$ that gives mass to the $Z'$ has $x_\Phi = x_1 - x_2$, so that dark Yukawa couplings are allowed. In this case the renormalizable dark sector Lagrangian reads
\begin{equation}\label{eq:Lagr}
- \mathcal{L} = m_1 \overline{\psi}_{1L} \psi_{1R} + m_2 \overline{\psi}_{2L} \psi_{2R} + y_1 \overline{\psi}_{1L} \Phi \psi_{2R} + y_2 \overline{\psi}_{2L} \Phi^\ast \psi_{1R} + \mathrm{h.c.}\,.
\end{equation}
In general, one physical phase cannot be removed from this Lagrangian. The dark quark mass matrix reads
\begin{equation}
\bmpsi = \begin{pmatrix} m_1 & y_1 v_\Phi \\ y_2 v_\Phi  & m_2 \end{pmatrix}\,.
\end{equation}
A motivated and simple limit is $y_2 \to 0$, which leads to $CP$ conservation (hence $\mathrm{Tr}(\sigma_2 \bm{X}^\prime_A) = \mathrm{Tr}(\sigma_2 \bm{X}^\prime_V) = \mathrm{Tr}[i\sigma_{1,3}(\bm{\zeta}^\prime - \bm{\zeta}^{\prime\,\dagger})] = 0$) and also has the advantage that compact analytical expressions can be derived for the relevant traces. In addition, we assume the further simplification $m_2 \to 0\,$: in this case one dark quark is massless, $\mathrm{Tr} (\bm{m}_{\psi^\prime}) = (m_1^2 + y_1^2 v_\Phi^2)^{1/2}$ and we find
\begin{align}
\mathrm{Tr}&(\sigma_1 \bm{X}^\prime_{A,V}) = -  \frac{ m_1 y_1 v_\Phi  }{m_1^2 + y_1^2 v_\Phi^2} (x_1 - x_2) \,,\qquad \mathrm{Tr}(\sigma_3 \bm{X}^\prime_{A,V}) =  \frac{ \{ -\,y_1^2 v_\Phi^2, m_1^2 \}}{ m_1^2 + y_1^2 v_\Phi^2   }(x_1 - x_2)\,, \;  \nonumber \\  
&\mathrm{Tr}[i\sigma_2 (\bm{\zeta}^\prime - \bm{\zeta}^{\prime\,\dagger})] =  - 2 y_1   \frac{m_1}{ ( m_1^2 + y_1^2 v_\Phi^2  )^{1/2} } \,, \qquad \mathrm{Tr}(\sigma_0\bm{X}^\prime_V) = x_1 + x_2\,. \label{eq:m2_0}
\end{align}
We see that for $y_1 v_\Phi/m_1 \sim O(1)$ both $CP$-odd dark pions $\hat{\pi}_{1,3}$ decay with $\mathrm{Tr}(\sigma_{1,3} \bm{X}^\prime_A) \sim 1$, whereas the $CP$-even dark pion $\hat{\pi}_2$ decays with $\mathrm{Tr}[i\sigma_2 (\bm{\zeta}^\prime - \bm{\zeta}^{\prime\,\dagger})] \sim y_1$.\footnote{In passing, we note that the same Lagrangian in Eq.~\eqref{eq:Lagr} can lead to a scenario where the dark pions do not decay through the $Z$ portal: assuming $CP$ conservation with $y_1 = y_2$ implies $U_L = U_R = U$, therefore $\bm{X}^\prime_A = 0$ while $\bm{X}^\prime_V = U^\dagger \bm{X} U$ is maximized.} 

A careful analysis shows that the dark pions mix with the angular and radial modes of $\Phi$, leading to corrections to the physical dark pion masses. These can be included in two equivalent ways: either in dark Chiral Perturbation Theory (ChPT), or by considering effective four-dark-quark operators. Both descriptions are provided for completeness in Appendix~\ref{app:dark_pion_mix}. Quantitatively, however, these effects are small and do not play a significant role in our discussion.

For the phenomenological study presented in Sec.~\ref{sec:pheno} we assume that $m_{\hat{\rho}} > 2 m_{\hat{\pi}}$ is always satisfied, hence the dark vector mesons decay rapidly to dark pions. In addition, we set $y_1 v_\Phi = m_1$ and $x_1 = - 1 = - x_2$, resulting in $\mathrm{Tr}(\sigma_{1,3} \bm{X}^\prime_A ) = 1$ and $ \mathrm{Tr}[ i\sigma_2 (\bm{\zeta}^\prime - \bm{\zeta}^{\prime\,\dagger}) ] = - \sqrt{2} \,y_1$.\footnote{An exact coincidence of $y_1 v_\Phi$ and $m_1$ is not motivated; we consider this limit purely as a simplification.} For definiteness, we also set the $U(1)'$ gauge coupling to $g_D = 0.25$. Thus $\hat{\pi}_{1,3}$ have identical effective decay constants
\begin{equation}
f_a^{(1,3)} = f_a \approx 3.9\;\mathrm{PeV} \, \left( \frac{1\;\mathrm{GeV}}{f_{\hat{\pi}}} \right) \bigg( \frac{10^{-2}}{\delta \hat{M}^2/ \hat{M}_{Z}^2}\bigg)  \left( \frac{ M_{Z^\prime}}{60 \text{ GeV}} \right)^2\,,
\end{equation}
hence identical lifetimes. On the other hand, $\hat{\pi}_2$ decays through Higgs mixing of strength
\begin{equation}
s_\theta^{(2)} \approx -\, 2.0 \times 10^{-8} \left( \frac{f_{\hat{\pi}}}{1\;\mathrm{GeV}} \right) \bigg(\frac{\kappa}{0.1} \bigg) \bigg( \frac{200\;\mathrm{GeV}}{m_\phi} \bigg)^2 \bigg( \frac{m_{\hat{\pi}}}{1\;\mathrm{GeV}} \bigg)^2\,,
\end{equation}
where $m^2_{\hat{\pi}} = 4 \sqrt{2} \pi f_{\hat{\pi}} y_1 v_\Phi$ has been used. Thus, the $\hat{\pi}_2$ is assumed to be effectively stable on collider scales in light of the much longer lifetime for any plausible values of $\kappa$ and $m_\phi$ (see the right panel of Fig.~\ref{fig:rho_BRs}).

\section{Dark pion phenomenology}\label{sec:pheno}

We are now in a position to discuss the phenomenology of dark pions, mediated by the $Z$ and $Z^\prime$. For dark pions lighter than a few GeV, the two main production mechanisms are: (i) exotic FCNC decays of SM mesons (primarily $B$ and $K$), which are relevant for the LHC, $B$ factories, and beam dump experiments; (ii) DS signals initiated by $Z$ or $Z^\prime$ in LHC collisions. Among the possible dark pion decay modes we focus on $\hat{\pi} \to \mu^+ \mu^-$, for which CMS and LHCb have exquisite sensitivity at low masses~\cite{CMS:2021sch,Aaij:2020ikh}. Concretely we consider the region $m_{\hat{\pi}} < 2$~GeV, where the branching ratio of the $CP$-odd dark pions to dimuons is always sizeable ($> 0.02$~\cite{Cheng:2021kjg}) and, at the same time, a sufficiently large hierarchy exists between the $Z'$ and $\hat{\pi}$ masses to justify a perturbative DS description of $Z^\prime \to \overline{\psi} \psi$ events at the LHC, even when the new vector is as light as $M_{Z^\prime} \sim O(10)$~GeV.

To make our study of the phenomenology quantitative, we adopt the benchmark dark sector presented in Sec.~\ref{sec:benchmark}. In particular, the two $CP$-odd dark pions decay with identical lifetimes whereas the $CP$-even dark pion is effectively stable on collider scales. While this benchmark is oversimplified, we believe it nevertheless captures the key features of the model at hand. The discussion is organized as follows. First, we consider dark pion production via FCNC decays of SM mesons and the associated constraints. Second, we present our new bounds from DS production at CMS and LHCb. Third, we explain the interplay of the two probes, also taking into account the indirect constraints presented in Sec.~\ref{sec:EWPO} and Sec.~\ref{sec:other_bounds}.

\subsection{Production from FCNC decays of SM mesons}\label{sec:FCNC}

Dark mesons can be produced from FCNC transitions of SM mesons, mediated by one-loop diagrams involving the $W$ boson. We focus on down-type FCNC effects, which receive a large contribution from top loops. Specifically, we discuss $B\to K^{(\ast)} \hat{\pi},\;K^{(\ast)} \hat{\rho}$ decays, which dominate except for very light dark mesons. In the present model only triangle diagrams mediating $\bar{d}_j d_i Z^{(\prime)}$ flavor-changing interactions play a role, whereas box diagrams are absent. Making use of the general amplitudes provided in Ref.~\cite{He:2009rz} to evaluate the $\bar{d}_j d_i Z^{(\prime)}$ vertices and integrating out $Z$ and $Z'$ at tree level, we arrive at the following four-fermion Lagrangian
\begin{equation}\label{eq:FCNC_4fermion}
\mathcal{L}_{\rm eff}^{\rm{FCNC}} = -\,  g_D \hat{g}_Z \frac{\delta \hat{M}^2}{ M_{Z}^2 M_{Z^\prime}^2 \cos^2\chi} \frac{g^2 }{128 \pi^2} J_{D}^\mu \bar{d}_j \gamma_\mu P_L d_i  \sum_{q\, \in\, u, c, t} V_{q j}^\ast V_{q i} \mathcal{K}_q + \mathrm{h.c.} ~,
\end{equation}
where
\begin{equation}
\mathcal{K}_q \equiv x_q \log \frac{ \Lambda_{\rm UV}^2 } {M_W^2} + \frac{-7x_q + x_q^2}{2(1 - x_q)}  - \frac{4x_q - 2 x_q^2 + x_q^3}{(1 - x_q)^2} \log x_q\, , 
\end{equation}
with the definition $x_q\equiv m_q^2/M_W^2$. The contribution from $q = t$ dominates. The residual divergence appears because our treatment of the $Z\,$-$\,Z'$ mixing is not UV complete. The divergence would be removed if the second Higgs doublet $H^\prime$, which induces $\delta \hat{M}^2$, were included dynamically (see for example Ref.~\cite{Dror:2017nsg}). In that case $\Lambda_{\rm UV}$ is set by the mass of the charged Higgs scalar. Numerically, the dimensionless quantity $\mathcal{K}_t$ varies from $5.0$ to $16$ as $\Lambda_{\rm UV}$ is increased from 300~GeV to 1~TeV, signaling an important theoretical uncertainty.

It is now straightforward to derive from Eq.~\eqref{eq:FCNC_4fermion}, using standard techniques (see Appendix~\ref{app:FCNC}), the branching ratios for $B$ meson decays to dark pions
\begin{equation}
\mathrm{BR}(B^{+,0} \to \{K^+ \hat{\pi}_b, K^{\ast 0} \hat{\pi}_b \} ) \approx \{ 0.92 , 1.1 \} \times 10^{-8}\, \bigg( \frac{1\;\mathrm{PeV}}{f_a^{(b)}} \bigg)^2 \bigg( \frac{\mathcal{K}_t}{10} \bigg)^2 \big\{ \lambda_{BK \hat{\pi}}^{1/2}, \lambda_{B K^\ast \hat{\pi}}^{3/2} \big\} \,, \label{eq:B_K_pihat}
\end{equation}
where we have chosen $\mathcal{K}_t = 10$ as the reference value, corresponding to a UV cutoff $\Lambda_{\rm UV} \approx 500\;\mathrm{GeV}$. Notice the appearance of the effective decay constant $f_a$, which was expected. For comparison, we also report results for final states containing dark vector mesons,\footnote{The result in Eq.~\eqref{eq:B0_Kstar_rhohat} is obtained by evaluating for $m_{\hat{\rho}} = 2\;\mathrm{GeV}$ the squared matrix element in the second and third lines of Eq.~\eqref{eq:B0_Kstar_rhohat_width}, which does not have a simple power-law dependence on $\lambda_{B K^\ast \hat{\rho}}$.}
\begin{align}
\mathrm{BR}(B^{+} \to K^+ \hat{\rho}_p ) \approx&\;  1.2 \times 10^{-9}\, \bigg( \frac{\varepsilon_{Zp}/m_{\hat{\rho}}}{10^{-6} \;\mathrm{GeV}^{-1}} \bigg)^2 \bigg( \frac{\mathcal{K}_t}{10} \bigg)^2 \lambda_{B K \hat{\rho}}^{3/2} \,, \label{eq:B0_K_rhohat} \\
\mathrm{BR}(B^{0} \to K^{\ast 0} \hat{\rho}_p ) \approx&\; 1.9 \times 10^{-9}\, \bigg( \frac{\varepsilon_{Zp}/m_{\hat{\rho}}}{10^{-6} \;\mathrm{GeV}^{-1}} \bigg)^2 \bigg( \frac{\mathcal{K}_t}{10} \bigg)^2 \,.\label{eq:B0_Kstar_rhohat}
\end{align}
These are somewhat suppressed compared to Eq.~\eqref{eq:B_K_pihat} for the parameter values we consider, therefore we neglect them in our discussion of the phenomenology. 

We set constraints on the model utilizing searches for $\hat{\pi} \to \mu^+ \mu^-$ decays. For the CHARM beam-dump experiment~\cite{CHARM:1985anb} we make use of the results of Ref.~\cite{Dobrich:2018jyi}, where bounds on $\mathrm{BR}(B \to K a) \times \mathrm{BR}(a \to \mu^+ \mu^-)$ for a pseudoscalar $a$ were provided. We translate those to our $(m_{\hat{\pi}}, f_a)$ plane. For CMS we recast the Run 2 scouting search for displaced dimuons~\cite{CMS:2021sch} (see also Ref.~\cite{Born:2023vll}). We simulate signal events with \texttt{PYTHIA8}~\cite{Bierlich:2022pfr}, including both $B\to K \hat{\pi}$ and $K^\ast \hat{\pi}$ followed by $\hat{\pi} \to \mu^+ \mu^-$, and calculate the signal yields using efficiency tables provided by CMS in the supplementary material of Ref.~\cite{CMS:2021sch}. From the same source we take the expected background yields, binned in the transverse displacement of the dimuon vertex $l_{xy}$, which is limited to be smaller than $11$~cm in the CMS analysis. For LHCb we apply the results of the Run 2 search in Ref.~\cite{Aaij:2020ikh}, using \texttt{PYTHIA8} to simulate signal events and including both $K\hat{\pi}$ and $K^\ast \hat{\pi}$, applying the cuts for the displaced $X\to \mu^+ \mu^-$ decay selection, and comparing the resulting signal yields with the provided cross section limits (owing to the non-negligible lifetime of $B$ mesons, we use the limits that do not require $X$ to originate from the primary vertex)~\cite{Aaij:2020ikh}. In both the CMS and LHCb searches, the best sensitivity is achieved when $\tau_{\hat{\pi}}$ is in the window $[0.1,1]$~cm.

\begin{figure}
\centering
\includegraphics[width=0.495\textwidth]{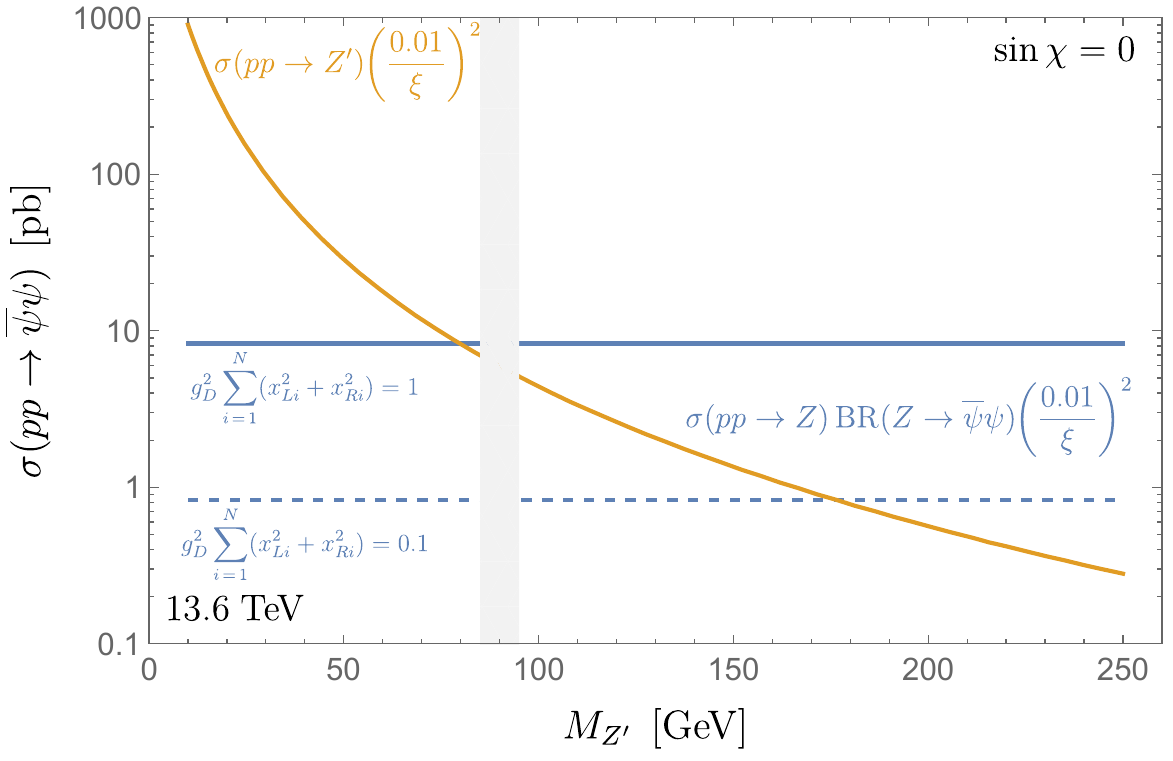}\hspace{0.2mm}
\includegraphics[width=0.495\textwidth]{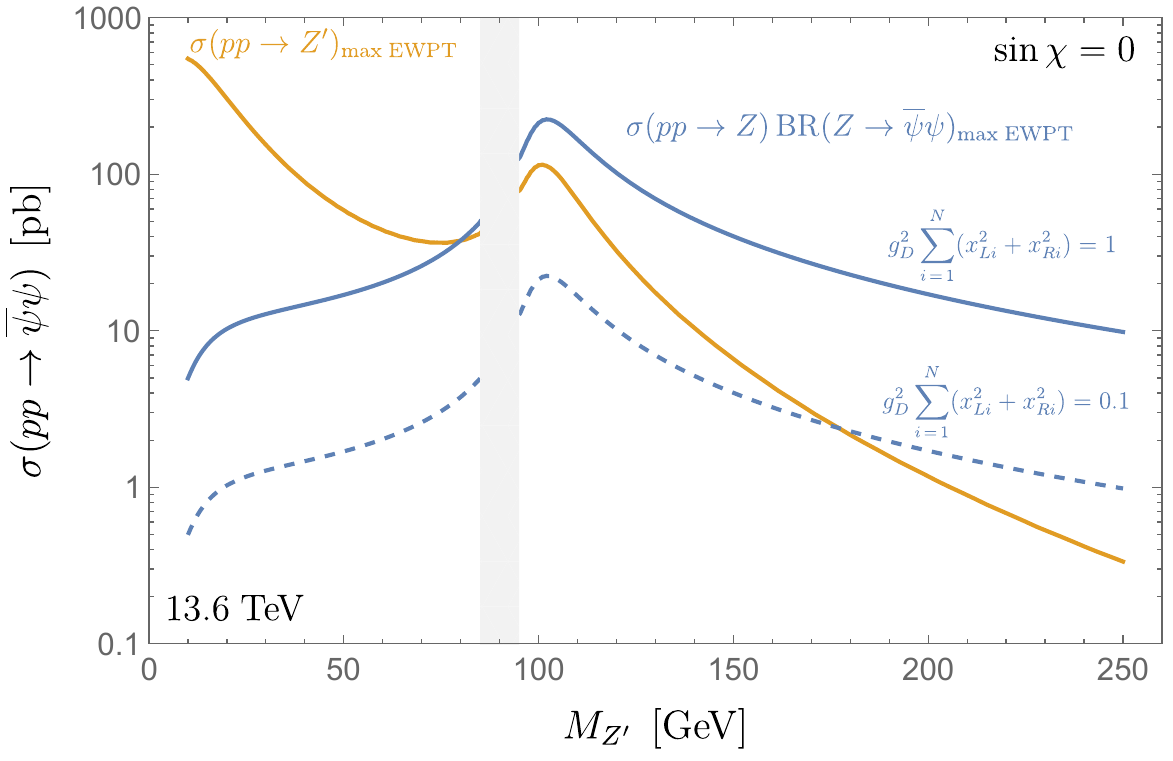}
\caption{Cross sections for production of dark quarks at the LHC, setting $\sin\chi = 0$. Orange curves show on-shell $Z^\prime$ production (the $Z^\prime f\bar{f}$ couplings are given by $\sin\xi$ times the $Z f\bar{f}$ couplings, and $\mathrm{BR}(Z^\prime \to \overline{\psi}\psi) \simeq 1$). Blue curves correspond to $Z$ production followed by decay to $\overline{\psi}\psi$, for two representative values of the sum over the dark quark $U(1)'$ charges appearing in $\mathrm{BR}(Z\to \overline{\psi} \psi)$. {\it Left panel:} a fixed mixing angle, $\xi = 0.01$, is taken as reference to show the production cross sections. {\it Right panel:} the largest production cross sections consistent with the $95\%$ CL upper bound on $\delta \hat{M}^2/\hat{M}_{Z}^2$ from EWPT (i.e.~the solid blue curve in Fig.~\ref{fig:benchmark_sinchi_0}) are shown.}
\label{fig:Z2_prod}
\end{figure}

\subsection{Production from dark showers in $Z$ or $Z^\prime$ decays \label{sec:Pheno}}
DS events at the LHC are initiated by the production of either a $Z$ or $Z^\prime$, which decays to a dark quark-antiquark pair. For the $Z$ the inclusive production cross section is $57.0\;$nb at $13.6\;$TeV,\footnote{The cross sections at $\{13, 14\}\,$TeV are $\{54.5, 58.9\}\,$nb.} and the branching ratio to dark quarks is obtained from Eq.~\eqref{eq:Z_to_psipsibar}. For the $Z'$, the cross section for on-shell production in $pp$ collisions at center of mass energy $\sqrt{s}$ reads
\begin{equation}
\sigma(pp \to Z^\prime) \simeq \frac{K_{Z^\prime} \pi \hat{g}_Z^2}{4N_c s} \sum_q (\bar{v}_q^2 + \bar{a}_q^2)\, L_{q\bar{q}} \left( \frac{M_{Z'}^2}{s} \right) \,,
\end{equation}
where the $\bar{v}_f, \bar{a}_f$ couplings are defined in Eq.~\eqref{eq:Z2_ff_couplings}, the factor $K_{Z^\prime} = 1.3$ approximately accounts for QCD corrections, and the parton luminosities are defined in Eq.~\eqref{eq:parton_lumi_def}. A light $Z'$ dominantly decays to dark quarks, with branching ratio very close to unity. The cross sections for $\overline{\psi}\psi$ production through either $Z$ or $Z'$ are shown in Fig.~\ref{fig:Z2_prod} as functions of $M_{Z^\prime}$, highlighting that for $M_{Z^\prime} \lesssim M_{Z}$ the decays of the $Z^\prime$ are the dominant DS production mechanism. For $M_{Z'}\lesssim 60$~GeV, EWPT constraints are compatible with a very large $Z'$ production rate at the LHC, corresponding to $O(10^7)$ events in the $O(100)$ fb$^{-1}$ of data collected by CMS in Run 2.

To set bounds on DS production from the already-discussed CMS~\cite{CMS:2021sch} and LHCb~\cite{Aaij:2020ikh} searches for low-mass displaced dimuon resonances, we separately simulate $pp \to Z$ or $Z' \to \overline{\psi}\psi$ events in \texttt{PYTHIA8}, approximately including initial state radiation of SM jets via the internal parton shower and treating the {\it dark} parton shower and hadronization using the Hidden Valley module~\cite{Carloni:2010tw,Carloni:2011kk,Albouy:2022cin}. For reference, the average dark pion multiplicities per $Z$ decay event are $13\,(4)$ for $m_{\hat{\pi}} = 0.3\,(2)$~GeV, whereas for a $20$~GeV $Z^\prime$ we find $7\,(3)$ as multiplicities for $m_{\hat{\pi}} = 0.3\,(2)$~GeV. We also account for the fact that in our benchmark model, on average, $2/3$ of the produced dark pions decay while $1/3$ remain invisible. Since we consider events with one displaced $\mu^+ \mu^-$ vertex, the selection and limit-setting procedures are similar to those we described in Sec.~\ref{sec:FCNC} for FCNC $B$ decays.

\subsection{Constraints and discussion}

\begin{figure}
\centering
\includegraphics[width=0.495\textwidth]{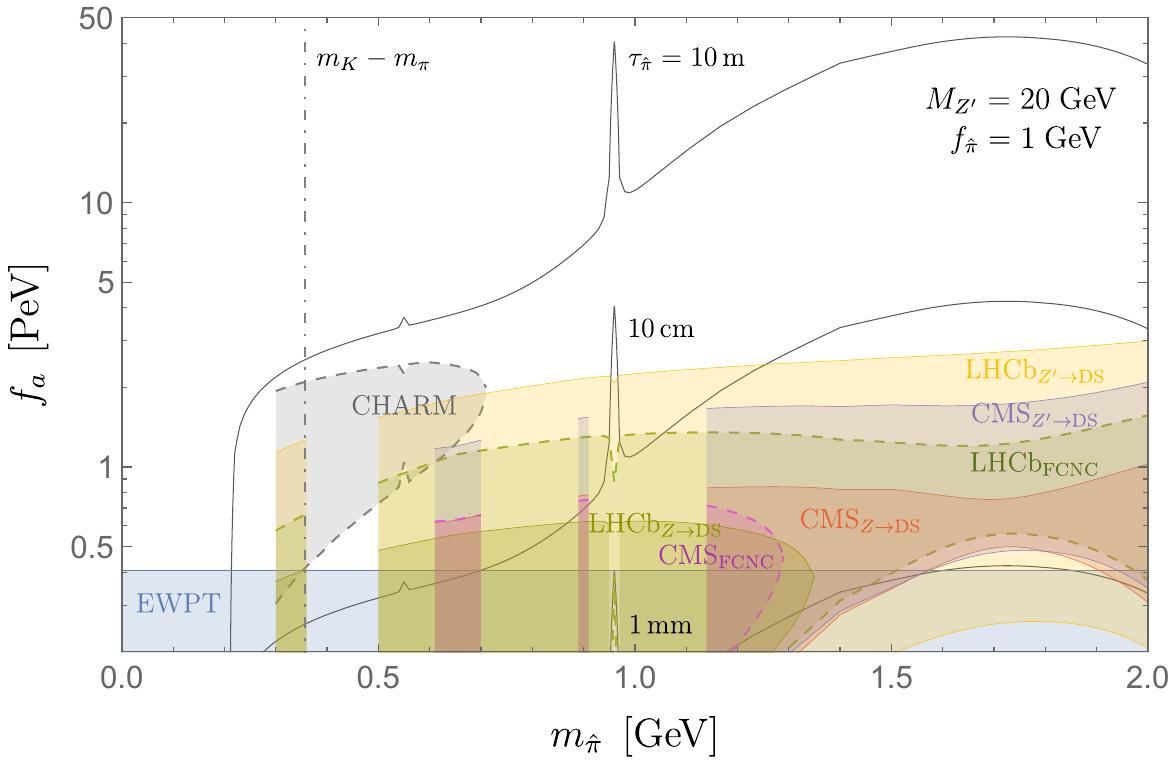}\vspace{3mm}
\includegraphics[width=0.495\textwidth]{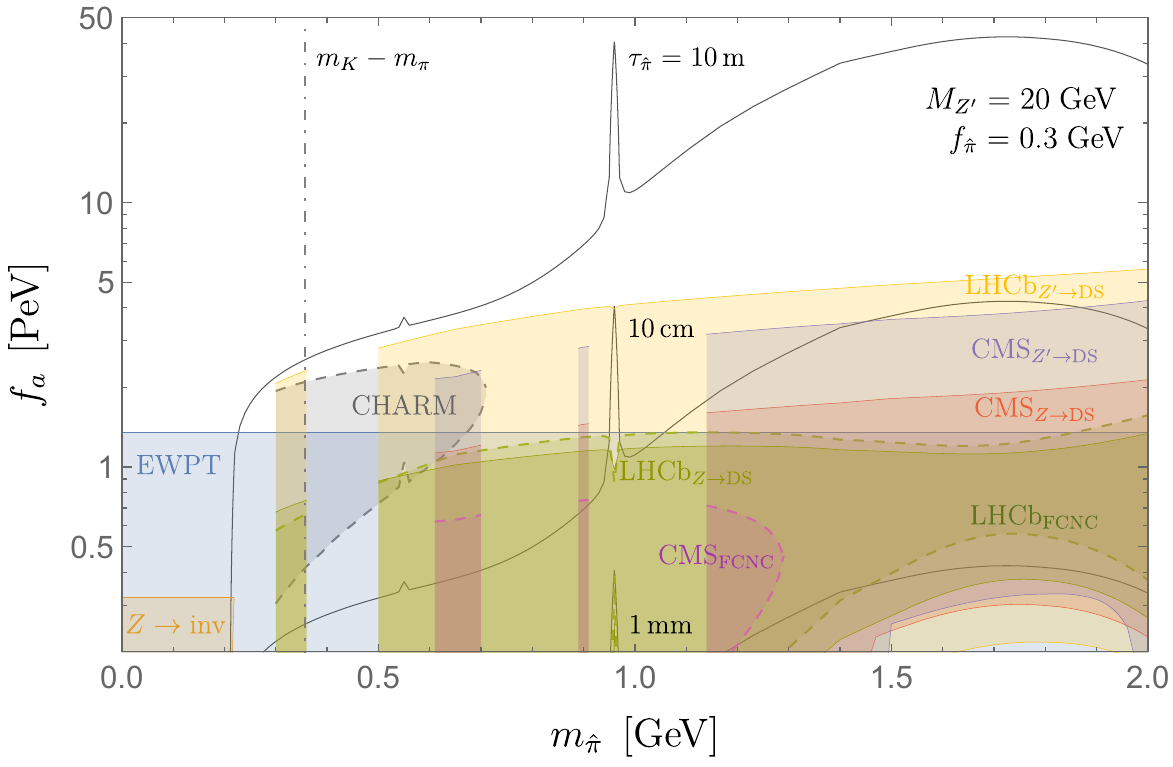}
\includegraphics[width=0.495\textwidth]{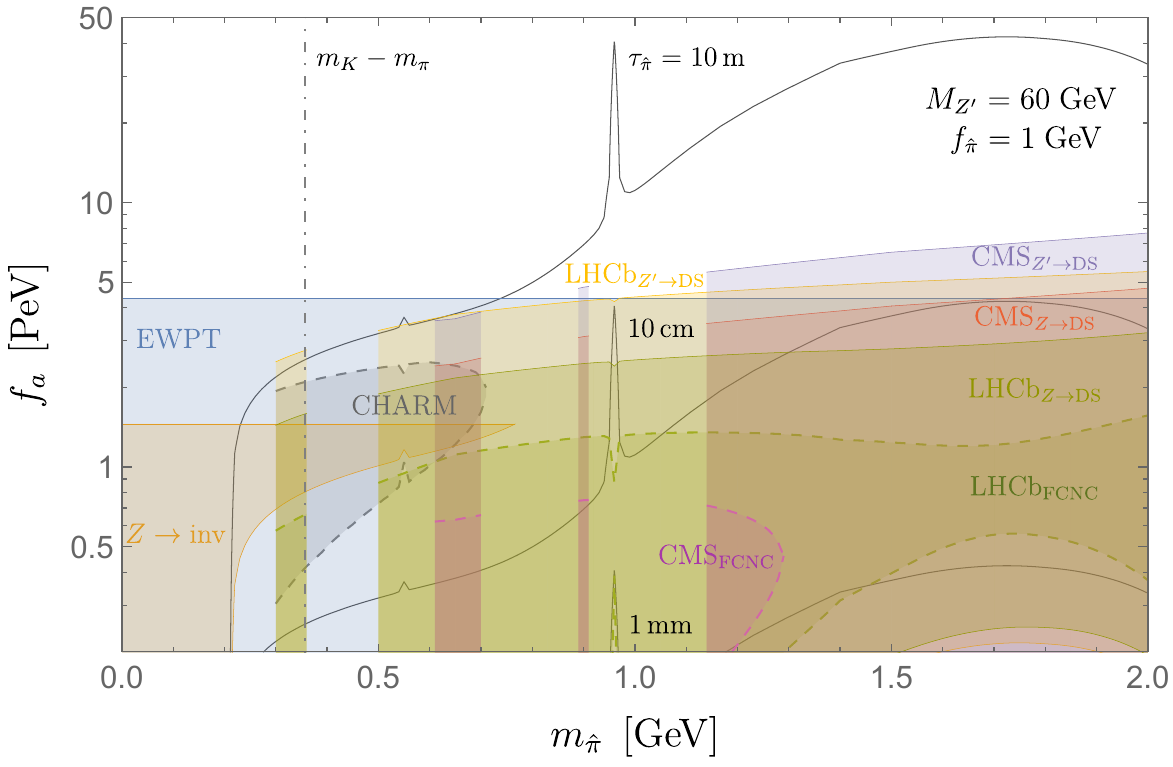}\vspace{3mm}
\includegraphics[width=0.495\textwidth]{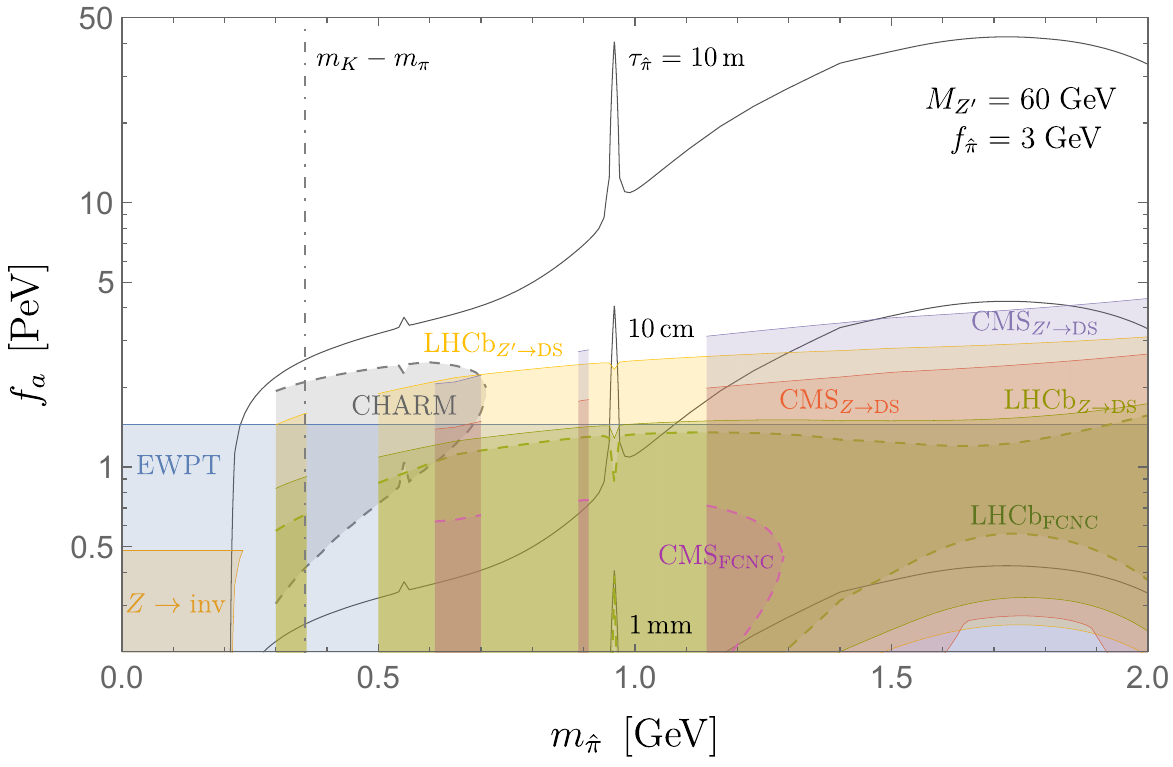}
\caption{Current constraints on the parameter space of our benchmark dark sector model with vanishing kinetic mixing. We take $M_{Z^\prime} = 20\;\mathrm{GeV}$ (top) or $M_{Z^\prime} = 60\;\mathrm{GeV}$ (bottom), and the indicated values of $f_{\hat{\pi}}$. Color-shaded regions with dashed boundaries show constraints from FCNC $B$ meson decays, while color-shaded regions with solid boundaries correspond to constraints from EWPT (blue), $Z\to \mathrm{invisible}$ (orange), $Z\to \mathrm{dark}\;\mathrm{showers}$ (red and green), and $Z^\prime \to \mathrm{dark}\;\mathrm{showers}$ (purple and yellow). The ``holes'' in the CMS~\cite{CMS:2021sch} and LHCb~\cite{Aaij:2020ikh} exclusion regions correspond to mass ranges that were masked in the experimental analyses, due to the proximity of SM resonances. The $Z\to \mathrm{invisible}$ constraint is too weak to appear in the top left panel. For $m_{\hat{\pi}} < m_K - m_\pi$ (i.e.,~to the left of the vertical dot-dashed line) bounds from kaon decays, not shown here, become relevant.}
\label{fig:paramspace_sinchi_0}
\end{figure}

Our main results are reported in Fig.~\ref{fig:paramspace_sinchi_0}, where we show the current constraints on the $(m_{\hat{\pi}}, f_a)$ plane for four different combinations of the mass of the light $Z^\prime$ and the dark pion decay constant $f_{\hat{\pi}}$. We assume the benchmark dark sector presented in Sec.~\ref{sec:benchmark} and, in addition, set the kinetic mixing to zero; taking $\sin\chi \neq 0$ would not change qualitatively the results, because the dark pion lifetimes would not be affected. For useful reference, we indicate with thin black curves three iso-contours of the ($CP$-odd) dark pion lifetime.

The blue-shaded regions indicate constraints from EWPT, which are insensitive to $m_{\hat{\pi}}$ and are translated by using the expression of the ALP effective decay constant,
\begin{equation}
f_a \propto \frac{ M_{Z'}^2 } {  f_{\hat{\pi}} \big( \delta \hat{M}^2 / \hat{M}_Z^2 \big) } \simeq \frac { M_{Z'}^2 } {  f_{\hat{\pi}}\, \xi }\,,
\end{equation}
where the last relation employs the approximate expression of the $Z\,$-$\,Z^\prime$ mixing angle for $\sin\chi = 0$ and $M_{Z'} \ll M_Z$ (exact formulae are used in the figure). We also show (orange shading) the regions ruled out by the $Z\to \mathrm{invisible}$ constraint, which we assume to hold only when the dark pion lifetimes are longer than $1$~m (changing this to, e.g., $10$~m would have a minor effect). We see that $Z\to \mathrm{invisible}$ plays a negligible role. In light of this we do not show the bound from $e^+ e^- \to \gamma Z'$ at DELPHI, which is similar to $Z\to \mathrm{invisible}$.

Colored regions with dashed boundaries are excluded by probes of FCNC $B$ meson decays at CHARM (gray shading), CMS (magenta) and LHCb (green). These depend directly on $f_a$ and are therefore identical in all panels, contrary to the other shown bounds, which are sensitive to the microscopic parameters of the theory and thus vary among the panels. CHARM excludes dark pion masses $m_{\hat{\pi}} \lesssim 700$~MeV for lifetimes between approximately $1$~m and $10$~m. LHCb excludes a significantly broader portion of parameter space than CMS does,\footnote{This conclusion partly depends on our conservative choice of including only $B \to K \hat{\pi}, K^\ast \hat{\pi}$ decays when setting the bounds for both CMS and LHCb. By contrast, in Ref.~\cite{Cheng:2021kjg} the CMS results~\cite{CMS:2021sch} were interpreted using an inclusive $B\to X_s \hat{\pi}$ branching ratio, assumed to equal $4\pm 1$ times the sum of $K \hat{\pi}$ and $K^\ast \hat{\pi}$. This resulted in comparable or even better sensitivity with respect to LHCb, for which the constraints shown in Ref.~\cite{Cheng:2021kjg} were based~\cite{Dobrich:2018jyi} on exclusive searches where the $K$ or $K^\ast$ are tagged~\cite{LHCb:2015nkv,LHCb:2016awg}, rather than the dedicated interpretation of a search requiring only a dimuon displaced vertex~\cite{Aaij:2020ikh} that we present here.} due to the unsuppressed $b\bar{b}$ production rate in the forward region, accompanied by a higher signal efficiency and smaller backgrounds. Yet, LHCb has probed new parameter space essentially only for the most favorable combination of $M_{Z'}$ and $f_{\hat{\pi}}$ considered (top left panel in Fig.~\ref{fig:paramspace_sinchi_0}), where the value of $f_a$ corresponding to the EWPT bound is smallest.

Colored regions with solid boundaries are excluded by our recasts of Run 2 CMS and LHCb searches for displaced dimuon resonances in terms of decays of the $Z$ or $Z^\prime$ to DS. We begin by discussing $Z\to \mathrm{DS}$ decays. CMS (exclusion region shown in red) reaches better sensitivity than LHCb~(green). However, the CMS search masked several $m(\mu^+\mu^-)$ windows corresponding to known resonances decaying to dimuons or a pair of charged hadrons ($K_S^0, \eta, \rho/\omega$, and $\phi$), which translate to important ``holes'' in our exclusion regions.\footnote{The window $300\;\mathrm{MeV} < m(\mu^+ \mu^-) \lesssim 400\;\mathrm{MeV}$ is not masked by CMS, but in this region of very small masses we do not trust our understanding of the efficiencies presented in the supplementary material of Ref.~\cite{CMS:2021sch}, hence we refrain from showing the corresponding constraints.} By contrast, LHCb only masked a region near the $K_S^0$ mass. We find that in most of the parameter space the $Z\to \mathrm{DS}$ signature has stronger sensitivity compared to FCNC $B$ meson decays, and improves (by a moderate amount) over the EWPT constraints. 

Finally, we turn to $Z^\prime \to \mathrm{DS}$ decays. Our recasts of the CMS~(purple) and LHCb~(yellow) dimuon resonance searches in terms of DS production by a light $Z^\prime$ are found to have the best sensitivity among all probes considered, extending well beyond EWPT constraints in all panels of Fig.~\ref{fig:paramspace_sinchi_0}. Furthermore, we observe a significant difference between the $M_{Z'} = 60$~GeV and $M_{Z'} = 20$~GeV cases: in the former the sensitivity of CMS is somewhat stronger, as already observed for $Z\to \mathrm{DS}$ decays; in the latter LHCb has a superior reach, because the production of a very light $Z'$ favors the forward region. In the $M_{Z'} = 20$~GeV scenario the LHCb sensitivity is impressive, improving over EWPT constraints by up to an order of magnitude in $f_a$.

\section{Conclusions\label{sec:Summary}}

The scenario where the $Z$ boson acts as a portal to a dark QCD sector presents striking opportunities to discover DS signals at the LHC. We presented one class of UV completions of the $Z$ portal, characterized by a dark $Z'$ that undergoes mass mixing with the $Z$. The dark $Z'$ can be light (since it is a SM singlet) and we focused especially on the possibility that $M_{Z'} < M_Z$, when the $Z'$ can contribute significantly to the DS production while remaining compatible with other data. 

To establish indirect constraints, we performed a careful analysis of EWPT for a dark $Z'$ with mass mixing and also kinetic mixing, which can be present in general. We pointed out that, in the presence of mass mixing, low-energy observables provide important complementary information to $Z$-pole measurements. Next, we discussed the EFT for the GeV-scale dark mesons, where both $Z$ and $Z'$ have been integrated out. The $CP$-odd dark pions, which can be viewed as composite ALPs, decay to SM fermions through the $Z\,$-$\,Z'$ mass mixing, with lifetimes falling between $1$~mm and $10$~m consistently with EWPT bounds. The $CP$-even dark pions decay through the Higgs-dark Higgs mixing, but are effectively stable on collider scales. 

Then, we recast CMS and LHCb searches for displaced dimuon resonances to DS signals originating from the on-shell production of $Z$ or $Z'$, where the visible signature is left by the decay to $\mu^+ \mu^-$ of any ($CP$-odd) dark pion in the DS. We found that, even though EWPT receive corrections at tree level, the CMS and LHCb sensitivity to DS is so strong that previously untouched parameter regions were tested in Run 2. Our recasts to DS signals were also compared to, and shown to have better sensitivity than, FCNC $B$ meson decays, where a single dark pion is produced at a time. This comparison can only be performed within a UV completion, since FCNC SM meson decays are sensitive to the ALP effective decay constant $f_a$, whereas DS signals depend separately on the parameters of the underlying model.

Together with Ref.~\cite{Cheng:2021kjg}, where the class of UV completions characterized by heavy fermions $Q$ were introduced, this paper completes the presentation of models that can generate the $Z$ portal at tree level. The phenomenological discussion was limited to current constraints. Projections to future LHC runs, including proposed auxiliary detectors, as well as estimates for future colliders will be presented in an upcoming companion work~\cite{Future:pheno}, where we examine a broad set of electroweak portals to a dark QCD sector. Model-specific limits will also be presented for both classes of UV completions of the $Z$ portal.

We conclude with a comment on dark matter in the $Z$ portal scenario, which has not been discussed thus far. The lightest dark baryon, stabilized by a dark baryon number symmetry, is a potential candidate for asymmetric dark matter. However, if it has a non-vanishing $U(1)^\prime$ charge it can scatter with ordinary matter through $Z$ (and $Z'$) exchange, leading to strong constraints from direct detection experiments. We present a brief discussion of dark baryon dark matter in Appendix~\ref{app:DM}. An alternative possibility, where some of the dark pions are stable and form dark matter, will be investigated in the future.

\acknowledgments

We thank Cristiano Alpigiani, Elias Bernreuther, Roberto Contino, Marco Ferla, Yee Bob Hsiung, Simon Knapen, Suchita Kulkarni, Jessie Shelton, Matthew Strassler, Indara Suarez, and Mia Tosi for useful discussions. We are also grateful to the authors of Refs.~\cite{Ilten:2018crw,Baruch:2022esd} for an exchange about DarkCast predictions. H.C. was supported by the Department of Energy under grant DE-SC0009999 and acknowledges the Aspen Center for Physics, which is supported by National Science Foundation grant PHY-2210452, where part of this work was performed. X.J. was supported in part by the National Natural Science Foundation of China under grant 12342502. L.L. was supported by NASA under grant 80NSSC22K081 and the Department of Energy under grant DE-SC0010010. E.S. was supported in part by the Science and Technology Facilities Council under the Ernest Rutherford Fellowship ST/X003612/1.

\clearpage

\appendix 
\section{Full Lagrangian and diagonalization of photon$\,$-$\,Z\,$-$\,$dark $Z'$ mixing}\label{app:details}
The model is described by the Lagrangian
\begin{align}
\mathcal{L} &= \mathcal{L}_{\rm SM} + \mathcal{L}_{\rm dark} + \mathcal {L}_{\rm mix},
\end{align}
with\footnote{We use $D_\mu = \partial_\mu + i g A_\mu$ as sign convention (note that this differs from Refs.~\cite{Cheng:2019yai,Cheng:2021kjg}). We partly adopt the notation of Ref.~\cite{Babu:1997st}.}
\begin{align}
\mathcal{L}_{\rm SM} & \supset -\frac{1}{4} \hat{B}_{\mu\nu} \hat{B}^{\mu\nu}  -\frac{1}{4} \hat{W}^3_{\mu\nu} \hat{W}^{3\mu\nu} + \frac{1}{2} \hat{M}_Z^2 \hat{Z}_\mu \hat{Z}^\mu  - \hat{e} \sum_f  \bar{f} \gamma^\mu \bigg( \frac{Y_f}{\hat{c}_W}  \hat{B}_\mu + \frac{T_{Lf}^3}{\hat{s}_W}  \hat{W}_\mu^3 \bigg) \hspace{-0.5mm}f , \\
 \mathcal{L}_{\rm dark} &= -\frac{1}{4} \hat{Z}'_{\mu\nu} \hat{Z}^{\prime \mu\nu} + (D_\mu \Phi)^\ast D^\mu \Phi - g_D \sum_{j\, =\, 1}^N  \left(\overline{\psi}_{j} \gamma^\mu x^j_{L} P_L \psi_{j} + \overline{\psi}_{j} \gamma^\mu x^j_{R} P_R \psi_{j}\right) \hat{Z}'_\mu  \label{eq:Zhat1}    \\
 - \frac{1}{4}&G^D_{a\,\mu\nu} G^{D\,\mu\nu}_a + \sum_{j\,=\,1}^N   i \overline{\psi}_j \slashed{D}_G \psi_j -  \sum_{i,\,j \,=\, 1}^N \Big( \overline{\psi}_{L i} m_{ij}  \psi_{R j} +  \overline{\psi}_{L i} \zeta^1_{ij} \psi_{Rj} \Phi + \overline{\psi}_{R i} \zeta^2_{ij} \psi_{Lj} \Phi  + \mathrm{h.c.} \Big) , \nonumber 
\end{align}
while $\mathcal {L}_{\rm mix}$ was given in Eq.~\eqref{eq:Zhat2}. Here $\hat{B}_{\mu\nu}$, $\hat{W}_{\mu\nu}^3$, $\hat{Z}'_{\mu\nu}$ are the field strengths of the $U(1)_Y$, third component of $SU(2)_L$, and $U(1)'$ respectively; $f = \{ u_L, \ldots\}$ are the SM chiral fermions and $Y_f$, $T_{Lf}^3$ are their charges under $U(1)_Y$ and the third generator of $SU(2)_L$. Furthermore, $\hat{Z}_\mu = \hat{c}_W \hat{W}_\mu^3 - \hat{s}_W \hat{B}_\mu$, $\hat{A}_\mu =\hat{s}_W \hat{W}_\mu^3 + \hat{c}_W \hat{B}_\mu$ are the $Z$ and photon fields without mixing with the new gauge boson. The dark quarks are written as $N$ Dirac fermions $\psi_{j}$, with $x^j_{L(R)}$ being the $U(1)'$ charges of their chiral components, and $D_G^\mu = \partial^\mu + i g_{s,\,D} \mathcal{T}^a G_D^a \,$, where $\mathcal{T}^a$ are the generators in the fundamental representation of $SU(N_d)$, is the dark QCD covariant derivative. The scalar field $\Phi$, singlet under the SM and with $U(1)^\prime$ charge $x_\Phi$ ($D_\mu = \partial_\mu + i g_D x_\Phi \hat{Z}^\prime_\mu$), is assumed to be dominantly responsible for the $\hat{Z}^\prime$ mass $\hat{M}_{Z^\prime}$ through spontaneous symmetry breaking. Depending on the $U(1)^\prime$ charge assignments, some (or all) entries of the dark Yukawa coupling matrices $\bm{\zeta}^{1,2}$ may be vanishing. 

The mass eigenstates of the neutral gauge bosons are obtained by first performing a field redefinition to remove the kinetic mixing,
\begin{equation}
\label{eq:kinetic_shift}
\begin{pmatrix} \hat{B}_\mu \\ \hat{Z}'_\mu \end{pmatrix} = \begin{pmatrix} 1 &\;\; -\tan \chi \\ 0 & \;\; 1/\cos\chi \end{pmatrix} \begin{pmatrix} \tilde{B}_\mu \\ \tilde{Z}'_\mu \end{pmatrix}.
\end{equation}
After this transformation, the mass matrix of $\tilde{Z}_\mu$-$\tilde{Z}'_\mu$ mixing, where $\tilde{Z}_\mu \equiv  \hat{c}_W \hat{W}_{\mu}^3 - \hat{s}_W \tilde{B}_\mu$, reads
\begin{equation}
\label{eq:mass_matrix}
\mathcal{M} \equiv \frac{1}{\cos^2 \chi} \begin{pmatrix}  
 \hat{M}_Z^2 \cos^2 \chi & \quad \delta \hat{M}^2 \cos \chi + \hat{M}_{Z}^2 \hat{s}_W \sin \chi \cos \chi \\ \delta \hat{M}^2 \cos \chi + \hat{M}_{Z}^2 \hat{s}_W \sin \chi \cos \chi & \quad \hat{M}_{Z'}^2 + \hat{M}_Z^2 \hat{s}_W^2 \sin^2 \chi + 2 \delta \hat{M}^2 \hat{s}_W \sin \chi \end{pmatrix}
 \end{equation}
 and is diagonalized to obtain the mass eigenstates $(Z_{\mu}, \, Z^\prime_{\mu})$ by
 \begin{equation}
 \begin{pmatrix} Z_{\mu} \\ Z^\prime_{\mu} \end{pmatrix} = \begin{pmatrix} \cos \xi &\; \sin \xi \\ -\sin \xi &\; \cos \xi \end{pmatrix} \begin{pmatrix} \tilde{Z}_\mu \\ \tilde{Z}'_\mu \end{pmatrix}\,,
 \end{equation} 
 with $\xi $ given in Eq.~\eqref{eq:mixing_angle}. Combining the above transformations we obtain the matrix $L$ in Eq.~\eqref{eq:transform}, as well as its inverse 
 \begin{align}
L^{-1} &\;= \begin{pmatrix} 1 &\;\; 0 &\;\; \hat{c}_W \sin \chi \\ 0 &\;\; \cos \xi &\;\; - \hat{s}_W \cos \xi \sin \chi + \sin \xi \cos \chi \\ 0 &\;\; -\sin \xi &\;\; \cos \xi  \cos \chi + \hat{s}_W \sin \xi \sin \chi \end{pmatrix}.
\end{align}
The physical masses of the $Z$ and $Z^\prime$ are
 \begin{equation}
 M^2_{Z, Z^\prime} = \frac{1}{2} \left( \mathcal{M}_{11} + \mathcal{M}_{22} \pm \mathrm{sgn} (\mathcal{M}_{11} - \mathcal{M}_{22}) \sqrt{(\mathcal{M}_{11} - \mathcal{M}_{22})^2 + 4 (\mathcal{M}_{12})^2}\,\right)\,.
 \end{equation}
In terms of $M_Z$, $M_{Z^\prime}$, and $\xi$, Eq.~\eqref{eq:mass_matrix} takes the form
\begin{equation}
\begin{pmatrix}
M_{Z}^2 \cos^2 \xi + M_{Z^\prime}^2 \sin^2 \xi & \quad (M_{Z}^2 - M_{Z^\prime}^2) \sin \xi \cos \xi \\ (M_{Z}^2 - M_{Z^\prime}^2) \sin \xi \cos \xi & \quad M_{Z}^2 \sin^2 \xi + M_{Z^\prime}^2 \cos^2 \xi
\end{pmatrix}\,,
\end{equation}
yielding in particular the exact relation
\begin{equation}
\hat{M}_Z^2 = M_{Z}^2 \left[ 1+ \sin^2 \xi \left( \frac{M_{Z^\prime}^2}{M_{Z}^2} -1 \right) \right]
\end{equation}
which has been used to derive some of our results. In turn, $s_W$ and $\hat{s}_W$ can be related by using $s_W c_W M_{Z} = \hat{s}_W \hat{c}_W \hat{M}_Z$ from Eq.~\eqref{eq:G_F}. Up to second order in $\xi$ one finds
\begin{equation}
\hat{s}_W^2 \simeq s_W^2 \left[ 1- \frac{c_W^2}{c_W^2 -s_W^2} \xi^2 \left(  \frac{M_{Z^\prime}^2}{M_{Z}^2} -1 \right)  \right].
\end{equation}

\section{Scalar potential}\label{app:potential}
Given the three scalar fields $H \sim \mathbf{2}_{1/2, \,0}\,$, $H^\prime \sim \mathbf{2}_{1/2, \,x_\Phi}\,$, $\Phi \sim \mathbf{1}_{0, \,x_\Phi}\,$ under $SU(2)_L \times U(1)_Y \times U(1)^\prime$, the most general renormalizable scalar potential is
\begin{align}
V =&\, - m_H^2 H^\dagger H + \lambda_H (H^\dagger H)^2 + m_{H^\prime}^2 H^{\prime \dagger} H^\prime + \lambda_{H^\prime} (H^{\prime \dagger} H^\prime)^2 - m_{\Phi}^2 \Phi^{\ast} \Phi + \lambda_{\Phi} (\Phi^\ast \Phi)^2 \nonumber \\
+&\, \lambda_{HH^\prime}^{(1)} (H^\dagger H) (H^{\prime \dagger} H^\prime) + \lambda_{HH^\prime}^{(2)} (H^\dagger H^\prime) (H^{\prime \dagger} H)   + \lambda_{H\Phi} H^\dagger H \Phi^\ast \Phi + \lambda_{H^\prime \Phi} H^{\prime \dagger} H^\prime \Phi^\ast \Phi \nonumber \\
-&\, \mu H^\dagger H^\prime \Phi^\ast + \mathrm{h.c.} , \label{eq:V}
\end{align}
where all the parameters except $\mu$ are real. The sign choices will become clear momentarily. The gauge boson masses have the expressions
\begin{equation}\label{eq:gauge_masses}
\hat{M}_Z^2 = \frac{\hat{g}_Z^2}{4}(v_H^2 + v_{H^\prime}^2 )\,, \quad \hat{M}_{Z'}^2 = g_D^2 (2 x_\Phi^2 v_\Phi^2 + x_{H^\prime}^2 v_{H'}^2 )\,, \quad \delta \hat{M}^2 = - \frac{1}{2} \hat{g}_Z g_D x_{H'} v_{H'}^2\,,
\end{equation}
where $x_{H'} = x_{\Phi}$.

We wish to demonstrate that this UV-complete model can produce at the same time: (i) a light $Z^\prime$ with $M_{Z^\prime} < M_Z\,$; (ii) a VEV for $H'$, with $v_{H'} \ll v_{\Phi}, v_H$, inducing a mass mixing between $Z$ and $Z'$ roughly as large as allowed by EWPT, $\delta \hat{M}^2/\hat{M}_Z^2 \lesssim 10^{-2}\,$; and (iii) scalar excitations of the second doublet $H^\prime$ that are heavy enough to justify neglecting their impact on the phenomenology below the weak scale, as we have done in this paper. Although a general analysis would be interesting, it is clearly beyond the present scope. Hence, we content ourselves with outlining how these features can be realized and presenting an explicit benchmark point for illustration.

The basic idea can be explained in a simpler way if the cross-quartic couplings in the second line of Eq.~\eqref{eq:V} are set to zero. We also choose a real $\mu > 0\,$. Going to the vacuum, $H^{(\prime)} = (0 \;\; h^{(\prime)} / \sqrt{2} )^T$ and $\Phi = \varphi$, we find then
\begin{equation}
V = - \frac{m_H^2}{2} h^2 + \frac{\lambda_H}{4} h^4 + \frac{m_{H^\prime}^2}{2} h^{\prime\, 2} + \frac{\lambda_{H^\prime}}{4} h^{\prime\, 4} - m_\Phi^2 \varphi^2 + \lambda_\Phi \varphi^4 - \mu h h^\prime \varphi\,.
\end{equation}
As hinted by the choice of signs in the quadratic terms, we assume that $H$ and $\Phi$ undergo spontaneous symmetry breaking even in the limit of vanishing trilinear coupling, $\mu \to 0$, whereas the VEV of $H^\prime$ is induced by $\mu$ and is given to a good approximation by $v_{H^\prime} \simeq \mu v_H v_\Phi/m_{H'}^2$. Before scalar mixing, the Higgs squared mass is $\tilde{m}_h^2 =  2 \lambda_H v_H^2 + \mu v_{H'} v_{\Phi} / v_H$, the radial mode of the SM-singlet $\Phi$ has $\tilde{m}_\phi^2 = 4 \lambda_\Phi v_\Phi^2 + \mu v_H v_{H'} / (2 v_\Phi)$, while $\tilde{m}_{h^\prime}^2 =  2 \lambda_{H^\prime} v_{H^\prime}^2 + \mu v_{H} v_{\Phi} / v_{H^\prime}$. 

We choose the benchmark point
\begin{align}
m^2_{H} =\, (82\;&\mathrm{GeV})^2\,,\quad \lambda_H \approx 0.135\,,\qquad m_\Phi^2  = (140\;\mathrm{GeV})^2\,, \quad \lambda_\Phi = 0.3 \,, \nonumber  \\
&m^2_{H'} = (500\;\mathrm{GeV})^2\,, \quad \lambda_{H'} = 0.3 \,, \qquad \mu = 100\;\mathrm{GeV}\,,
\end{align}
which leads to the VEVs $\{v_H, v_{\Phi}, v_{H'} \} \approx \{ 245.6, 186.2, 18.3\}$~GeV, hence $v = ( v_H^2 + v_{H'}^2 )^{1/2} \approx 246.2$~GeV. Notice that the analytical expression of $v_{H'}$ as induced by the $\mu$ tadpole is indeed accurate, $v_{H^\prime} = \mu v_H v_\Phi/m_{H'}^2 \approx 18.3\;\mathrm{GeV}$. After accounting for mixing, the physical masses of the $CP$-even scalars are $m_h \approx 126$~GeV, $m_\phi \approx 204$~GeV, and $m_{h'} \approx 503$~GeV; in particular, the state mostly composed of the radial mode of $\Phi$ is heavier than the Higgs. As an example we fix the $Z'$ mass, or more precisely $\hat{M}_{Z'} = g_D |x_{H'}| ( 2v_\Phi^2 + v_{H'}^2 )^{1/2}$, to $60$~GeV. We thus obtain $g_D \approx 0.11$, where we have taken $x_{H'} = -2$ as in our benchmark example. Then from Eq.~\eqref{eq:gauge_masses} we find the mass mixing parameter $\log_{10} (\delta \hat{M}^2 / \hat{M}^2_{Z} ) \approx -2.5$, which is not far from the current upper bound from EWPT presented in the right panel of Fig.~\ref{fig:benchmarks}. The masses of all the physical scalars contained (mostly) in $H'$ are approximately set by $(m_{H'}^2)^{1/2} = 500\;\mathrm{GeV}$, which we expect to be large enough to make them compatible with existing constraints. 

Integrating out the $H'$ doublet from Eq.~\eqref{eq:V} leads to a low-energy quartic coupling $\kappa = \lambda_{H\Phi} - |\mu|^2 / m_{H'}^2$. For the benchmark point we have chosen, $\lambda_{H\Phi} = 0$. Hence $\kappa \approx -\, 0.04$ and the $h\,$-$\,\phi$ mixing angle is given approximately by
\begin{equation}
\theta_s \simeq \frac{\sqrt{2}\, \mu^2 v_\Phi v}{m_{H'}^2 (\tilde{m}_\phi^2 - \tilde{m}_h^2)} \approx 0.10\,.
\end{equation}
Hence the single-Higgs couplings are suppressed by $- \theta_s^2/2 \approx -\,0.5\%$ relative to their SM values, an effect which lies beyond the reach of the (HL)-LHC, but can be tested at future colliders~\cite{deBlas:2019rxi}.

 \section{More on electroweak precision constraints}
 \label{app:EW_heavyZprime}
In this Appendix we present further details on our calculation of EWPT. Making use of the results in Appendix~\ref{app:details} we derive the interaction of the on-shell $Z$ with SM fermions,\footnote{The result differs from Ref.~\cite{Babu:1997st}, due to the appearance of $M_{Z^\prime}^2/M_{Z}^2 - 2$ in our $O(\xi^2)$ term instead of $M_{Z^\prime}^2/M_{Z}^2 -1$. Taking the latter expression would yield incorrect results for $S, T, U$ at $O(\xi^2)$.}
\begin{equation}\label{eq:Z1ff_model}
\mathcal{L}_{Z f\bar{f}}= -\frac{e}{s_W c_W}  \bar{f} \gamma^\mu \left[ 1 + s_W \xi   \tan \chi +  \frac{\xi^2}{2} \bigg( \frac{M_{Z^\prime}^2}{M_{Z}^2} -2\bigg) \right] ( T_{Lf}^3 - s_\ast^2 Q_f )   f Z_{\mu} ,
\end{equation}
where $s_\ast^2 \equiv \sin^2 \theta_\ast$ is given by
\begin{equation}\label{eq:s_star}
\frac{s_\ast^2}{s_W^2} = 1 + \frac{c_W^2}{s_W}  \xi \tan \chi  - \frac{c_W^2}{c_W^2- s_W^2} \xi^2 \bigg( \frac{M_{Z^\prime}^2}{M_{Z}^2} -1\bigg) \, ,
\end{equation}
while the $W$ mass is
\begin{equation}\label{eq:MW2model}
\frac{M_W^2}{M_{Z}^2} = c_W^2 \left[ 1 + \frac{c_W^2}{c_W^2 - s_W^2} \xi^2 \left( \frac{M_{Z^\prime}^2}{M_{Z}^2} - 1 \right) \right]\,.
\end{equation}
Equations~\eqref{eq:Z1ff_model},~\eqref{eq:s_star} and~\eqref{eq:MW2model} lead to the identifications of $S,T,U$ in Eq.~\eqref{eq:PeskinTakeuchi}. The experimental values are given by~\cite{ParticleDataGroup:2022pth}
\begin{align} \label{eq:STU_exp}
S = - 0.02 \pm 0.10\,,\qquad T = 0.03 \pm 0.12\,, \qquad U = 0.01 \pm 0.11\,, 
\end{align}
with
\begin{equation}
\rho = \begin{pmatrix} 1 & 0.92 & -0.80 \\ & 1 & -0.93 \\ & & 1 \end{pmatrix}\,
\end{equation}
as the correlation matrix.

The weak charge of the atom is defined as
\begin{equation}\label{eq:QW}
Q_W = -2 \rho_\ast (0) \left[ Z \Big( - \frac{1}{2} + 2 s_\ast^2 (0) \Big) +  \frac{N}{2} \right]\,.
\end{equation}
Its expression including BSM corrections,
\begin{equation}
Q_W \approx Q_W^{\rm SM} \left( 1 +  \frac{\delta Q_W}{Q_W} \right)\,,
\end{equation}
gives for Caesium-133 ($Z = 55$ and $N = 78$), using $\rho_\ast (0)_{\rm SM} \approx 1$ and $s^2_\ast (2.4\;\mathrm{MeV})_{\rm SM} = 0.2386$~\cite{ParticleDataGroup:2022pth}, the result in Eq.~\eqref{eq:QW_prediction}. The SM prediction is $Q_W^{\rm SM} = -73.24$ (with negligible uncertainty), whereas the experimental value is $Q_W^{\rm exp} = -72.82 \pm 0.42$~\cite{ParticleDataGroup:2022pth}.

For the E158 measurement of parity violation in $e^- e^- \to e^- e^-$ scattering at low energy~\cite{SLACE158:2005uay}, $\rho_\ast (0)_{\rm SM} \approx 1$ and $s^2_\ast (161\;\mathrm{MeV})_{\rm SM} = 0.2385$~\cite{ParticleDataGroup:2022pth} were used to obtain the second expression in Eq.~\eqref{eq:ee_ee}. The SM prediction is $g_{AV}^{ee,\,\mathrm{SM}} = 0.0226$, and the experimental value is $g_{AV}^{ee,\,\mathrm{exp}} = 0.0190 \pm 0.0027$~\cite{Erler:2013xha}.

For the $Q_{\rm weak}$ measurement of the right-left asymmetry in $e^- p \to e^- p$ scattering~\cite{Qweak:2018tjf}, $\rho_\ast (0)_{\rm SM} \approx 1$ and $s^2_\ast (157\;\mathrm{MeV})_{\rm SM} = 0.2385$~\cite{ParticleDataGroup:2022pth} were used to derive the second expression in Eq.~\eqref{eq:ep_ep}. The SM prediction is $g_{AV}^{ep,\,\mathrm{SM}} = - 0.0355$, and the experimental value is $g_{AV}^{ep,\,\mathrm{exp}} = -0.0356 \pm 0.0023$~\cite{ParticleDataGroup:2022pth}.

\begin{figure}
\centering
\includegraphics[width=0.5\textwidth]{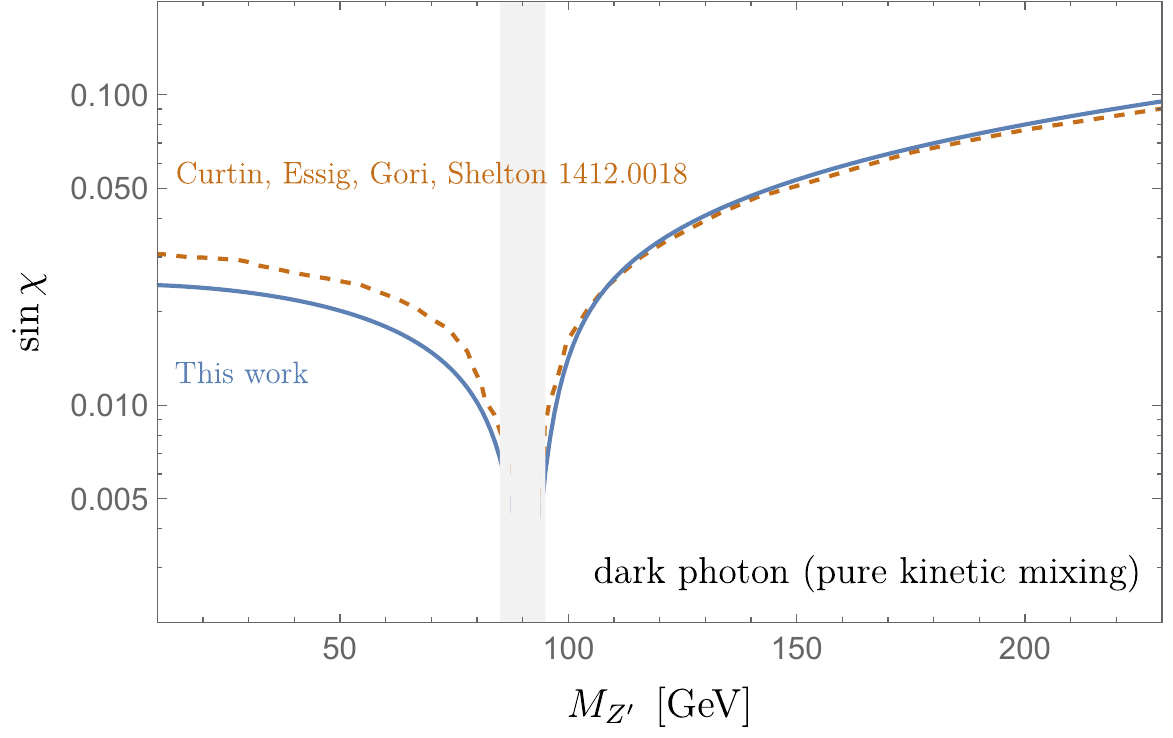}
\caption{$95\%$ CL constraints from electroweak precision data on the kinetic mixing parameter $\sin\chi$, assuming the mass mixing vanishes ($\delta \hat{M}^2 = 0$). The solid blue curve shows our result. The dashed brown line is taken from Ref.~\cite{Curtin:2014cca}, accounting for the different normalization (namely, their $\varepsilon$ corresponds to our $c_W \sin\chi$).}
\label{fig:DP_comparison}
\end{figure}

To set electroweak precision constraints we form a $\chi^2$ that combines both $Z$-pole and low-energy observables,
\begin{equation}\label{eq:total_chi2}
\chi^2 = \chi^2_{STU} + \chi^2_{\rm low}\,,
\end{equation}
where
\begin{align}
\chi^2_{STU} \,=&\, \sum_{i,\,j \,=\, 1}^3 (x_i - x_i^{\rm exp}) (C^{-1})_{ij} (x_j - x_j^{\rm exp})\,,\qquad C_{ij} = \sigma_i \rho_{ij} \sigma_j\,,
\end{align}
is the $Z$-pole piece, where the $x_i$ are defined by writing Eq.~\eqref{eq:STU_exp} as $\vec{x} = \vec{x}^{\,\rm exp} \pm \vec{\sigma}$, and
\begin{equation}
\chi^2_{\rm low} = \left( \frac{Q_W - Q_W^{\rm exp}}{\sigma_{Q_W}} \right)^2 +  \bigg( \frac{g_{AV}^{ee} - g_{AV}^{ee,\,\mathrm{exp}}}{\sigma_{g_{AV}^{ee}}} \bigg)^2 +  \bigg( \frac{g_{AV}^{ep} - g_{AV}^{ep,\,\mathrm{exp}}}{\sigma_{g_{AV}^{ep}}} \bigg)^2
\end{equation}
is the low-energy piece. For useful reference, in Fig.~\ref{fig:DP_comparison} we compare the bound obtained from our fit for pure kinetic mixing (in this case, the impact of low-energy observables in Eq.~\eqref{eq:total_chi2} is negligible), to the widely-used result of Ref.~\cite{Curtin:2014cca}. We find excellent agreement for $M_{Z'} > M_Z$ and consistency within $20\%$ for $M_{Z'} < M_Z$.

\subsection{Oblique parameters for heavy dark $Z'$}
For $M_{Z^\prime} \gg M_{Z}$, the dark $Z'$ can be integrated out to obtain an effective Lagrangian at the weak scale. In the parametrization of Ref.~\cite{Barbieri:2004qk}, Eqs.~\eqref{eq:s2star} and~\eqref{eq:MW2} are replaced by expressions that depend on $W, Y, V, X$ in addition to $\widehat{S}, \widehat{T}$ and $\widehat{U}$,
\begin{align}
\bar{Z} \,=&\, 1 + \frac{1}{2} \left[ \widehat{T} - W - \frac{s_W^2}{c_W^2} Y + 2 \frac{s_W}{c_W} X \right]\,, \label{eq:Barbieri_version} \\
s_\ast^2 (M_Z^2) - s_W^2 \,=&\, \frac{1}{c_W^2 - s_W^2} \left[ s_W^2 \widehat{S} - s_W^2 c_W^2 \widehat{T}  - s_W^4 W + \frac{s_W}{c_W} ( 1 - 2 s_W^2 c_W^2) X - s_W^2 c_W^2 Y \right]\,, \nonumber \\
\frac{M_W^2}{M_Z^2} - c_W^2 \,=&\, \frac{c_W^2}{c_W^2 - s_W^2} \left[ - 2 s_W^2 \widehat{S} + c_W^2 \widehat{T} - (c_W^2 - s_W^2) ( \widehat{U} - V ) - 2 s_W c_W X + s_W^2 (W + Y ) \right]\,. \nonumber
\end{align}
Integrating out the $\hat{Z}^\prime$ at tree level from Eqs.~\eqref{eq:Zhat1} and~\eqref{eq:Zhat2}, we obtain
\begin{align}
\widehat{S} \,=&\, - \frac{\hat{c}_W^2}{\hat{s}_W} \frac{\delta \hat{M}^2}{\hat{M}_{Z'}^2} \bigg( \sin\chi + \hat{s}_W \frac{ \delta \hat{M}^2 }{\hat{M}_{Z'}^2 } \bigg)\,,\qquad
\widehat{T} \,=\, \hat{c}_W^2 \frac{ ( \delta \hat{M}^2)^2}{M_W^2 \hat{M}_{Z'}^2} \,, \nonumber \\
\widehat{U} \,=&\, - \hat{c}_W^2 \frac{(\delta \hat{M}^2)^2}{\hat{M}^4_{Z'}}\,, \quad W = V = \hat{c}_W^2 \frac{M_W^2 (\delta \hat{M}^2)^2}{\hat{M}_{Z'}^6}\,, \quad Y = \frac{M_W^2}{\hat{M}_{Z'}^2} \bigg( \sin \chi + \hat{s}_W \frac{ \delta \hat{M}^2 }{\hat{M}_{Z'}^2 } \bigg)^2\,, \nonumber \\ 
X \,=&\, - \hat{c}_W \frac{M_W^2 \delta \hat{M}^2}{\hat{M}_{Z'}^4} \bigg(  \sin \chi + \hat{s}_W \frac{ \delta \hat{M}^2 }{ \hat{M}_{Z'}^2} \bigg)\,. \label{eq:Barbieri_params}
\end{align}
We observe that $\widehat{S} = (\hat{c}_W / \hat{s}_W) ( \hat{M}_{Z'}^2 / M_W^2 ) X$ and $\widehat{T} = - (\hat{M}_{Z'}^2/M_W^2)\, \widehat{U} = (\hat{M}_{Z'}^4 / M_W^4) V$, explicitly illustrating the well-known fact that $\widehat{S}, \widehat{T}, W, Y$ are sufficient to describe heavy universal new physics~\cite{Barbieri:2004qk}. We also find that the inequality $WY \geq X^2$~\cite{McCullough:2023szd} is saturated. For pure mass mixing, only $\widehat{T}$ is generated at dimension-$6$ level while the other parameters arise at higher dimensions.

It is straightforward to check that Eqs.~\eqref{eq:Barbieri_version} and~\eqref{eq:Barbieri_params} agree with Eqs.~(\ref{eq:s2star}$\,$--$\,$\ref{eq:PeskinTakeuchi}), once the latter are expanded up to quadratic order in $\delta \hat{M}^2/\hat{M}_Z^2$ or $\sin\chi$. Nevertheless, the individual oblique parameters take up {\it genuinely different} values in the two descriptions. For instance, for pure kinetic mixing $\delta \hat{M}^2 = 0$ we immediately find from Eq.~\eqref{eq:Barbieri_params} that only $Y = \sin^2\chi M_W^2 /\hat{M}_{Z'}^2$ is non-vanishing, whereas from Eq.~\eqref{eq:PeskinTakeuchi} we obtain $c_W^2 \alpha T \simeq \alpha S/4 \simeq - s_W^2 \sin^2 \chi\, M_W^2 / \hat{M}_{Z'}^2\,$, where $\sin \chi \ll 1$ was assumed.

\section{Further details on the phenomenology}\label{app:pheno_details}
In this Appendix we collect several results, which we hope will be helpful for phenomenological studies beyond the present paper. 
\subsection{Constraints on on-shell $Z'$ production}\label{app:onshell_Zprime}
To set constraints on the dark $Z^\prime$ from the CMS search for $\mu^+ \mu^-$ resonances~\cite{CMS:2019buh} we make use of the results for the standard dark photon model (i.e.,~with pure kinetic mixing and no decays to the dark sector). We require
\begin{align}\label{eq:bound_Zprime_dileptons}
\sum_q (\bar{v}_q^2 + \bar{a}_q^2) & L_{q\bar{q}} \left( \frac{M_{Z'}^2}{s} \right)  \frac{\Gamma (Z' \to \mu^+ \mu^-)}{\sum_f \Gamma (Z' \to \bar{f} f) + \Gamma(Z' \to \overline{\psi} \psi)} \\
& = \sum_q (\bar{v}_q^2 + \bar{a}_q^2) L_{q\bar{q}} \left( \frac{M_{Z'}^2}{s} \right) \frac{\Gamma (Z' \to \mu^+ \mu^-)}{\sum_f \Gamma (Z' \to \bar{f} f)}\Bigg|_{\delta \hat{M}^2\, =\, 0\,,\;\sin \chi \,=\, \varepsilon / c_W\,, \;M_{Z^\prime} \,=\, m_{A'} } \,,\nonumber
\end{align}
where the parton luminosities are defined as 
\begin{equation}\label{eq:parton_lumi_def}
L_{q\bar{q}} (\tau) = \int_{\tau}^{1} \frac{dx}{x} \big[ f_q (x) f_{\bar{q}} (\tau/x) + \{q \leftrightarrow \bar{q} \} \big]
\end{equation}
and evaluated using MSTW2008NLO PDFs~\cite{Martin:2009iq}, with factorization scale set to $M_{Z'}/2$. On the right-hand side of Eq.~\eqref{eq:bound_Zprime_dileptons}, the limit on the kinetic mixing parameter $\varepsilon$ as a function of $m_{A'}$ is taken from Ref.~\cite{CMS:2019buh}. Thus, for each value of $M_{Z'}$ and choice of $g_D^2 \sum_{i\, =\, 1}^N (x_{L i}^2 + x_{R i}^2 )$ we obtain a bound on the $\big( \delta \hat{M}^2 / \hat{M}_Z^2 , \sin \chi \big)$ parameter space. The constraints for $\sin\chi = 0$ are shown in the right panel of Fig.~\ref{fig:benchmark_sinchi_0}.

For the search for $e^+ e^- \to \gamma Z^\prime$ followed by $Z^\prime \to \mathrm{invisible}$ at LEP2~\cite{DELPHI:2003dlq,Fox:2011fx}, we reinterpret the results for a dark photon with $\mathrm{BR}(A'\to \mathrm{invisible}) \approx 1$ shown in Refs.~\cite{Ilten:2018crw,Hochberg:2015vrg}. We require 
\begin{equation}\label{eq:DELPHI_limit}
\bar{v}_e^2 + \bar{a}_e^2 = (\bar{v}_e^2 + \bar{a}_e^2 )\Big|_{\delta \hat{M}^2\, =\, 0\,,\;\sin \chi \,=\, \varepsilon / c_W\,, \;M_{Z^\prime} \,=\, m_{A'} }\,,
\end{equation}
where on the right-hand side the limit on $\varepsilon$ as a function of $m_{A'}$ is taken from Ref.~\cite{Ilten:2018crw} for $m_{A'} < M_Z$, whereas for $m_{A^\prime} > M_Z$ we apply the limit on $\varepsilon$ as given in Ref.~\cite{Hochberg:2015vrg}. The resulting bound for $\sin\chi = 0$ is given in the right panel of Fig.~\ref{fig:benchmark_sinchi_0}.

\subsection{Higgs decays to the dark sector mediated by scalar mixing}\label{app:h_decays}

The mixing of scalars induces a coupling between the 125 GeV Higgs boson and the dark quarks,
\begin{equation}
\mathcal{L} \supset  - \zeta_{ij} \sin \theta_s \overline{\psi}_{Li} \psi_{Rj} \frac{h}{\sqrt{2}} +  \text{h.c.}\,, 
\end{equation}
allowing the Higgs boson to decay into dark quarks, 
\begin{equation}
\Gamma ( h \to \overline{\psi} \psi) \simeq \frac{N_d m_h}{16 \pi} {\rm Tr} \big( \bm{\zeta}^\dagger \bm{\zeta} \big) \sin^2 \theta_s \, .
\end{equation}
If $m_\phi < m_h/2$, the Higgs boson can also decay to two $\phi$ (which subsequently decay to dark hadrons),
\begin{equation}
\Gamma ( h \to \phi \phi) \simeq \frac{\kappa^2 v^2}{32 \pi m_h} \bigg( 1- \frac{4 m_\phi^2}{m_h^2}\bigg)^{1/2}.
\end{equation}
If one assumes that most decays to the dark sector are invisible, then the current LHC bound $\mathrm{BR}(h \to \mathrm{invisible}) < 0.11$~\cite{ATLAS:2020kdi} translates to
\begin{equation}
\label{eq:hinv}
{\rm Tr} \big(\bm{\zeta}^\dagger \bm{\zeta}\big) \sin^2\theta_s < 7 \times  10^{-5}\,,\qquad
\text{and for } m_\phi \ll m_h/2, \quad  \kappa < 10^{-2} .
\end{equation}
Additionally, the Higgs can decay to two $Z'$ vectors if $M_{Z^\prime} < m_h/2$,
\begin{equation}
\Gamma(h \to Z^\prime Z^\prime) \simeq \frac{m_h^3 \sin^2 \theta_s }{64\pi v_\Phi^2} \bigg( 1 - 4\, \frac{M_{Z^\prime}^{2}}{m_h^2} + 12 \frac{M_{Z^\prime}^4}{m_h^4} \bigg) \bigg(1 - \frac{4M_{Z^\prime}^2}{m_h^2} \bigg)^{1/2}\,,
\end{equation}
where $M_{Z^\prime}^2 \approx \hat{M}_{Z'}^2 \simeq 2 g_D^2 x_\Phi^2 v_\Phi^2$ has been used. 

\subsection{$\hat{\pi}_a  \to \hat{\pi}_b\, f \bar{f}$ decays}\label{app:pia_pibff}

The $\hat{\pi}_a  \to \hat{\pi}_b\, f \bar{f}$ modes can be relevant if ${\rm Tr}(\sigma_a \bm{X}'_A)$ is suppressed and significant mass splittings exist between the dark pions. These decays are mediated by the vector current (instead of the axial vector current, which mediates direct decays of the $CP$-odd dark pions to the SM). They are treated similarly to semileptonic kaon decays in the SM, such as $K \to \pi\, \ell \nu$ (see e.g.~Ref.~\cite{Cirigliano:2011ny} for a review). Starting from the four-fermion Lagrangian Eq.~\eqref{eq:4f_SM_dark}, we make use of the hadronic matrix element
\begin{equation}
\langle \hat{\pi}_b (p_2) | J_{DV}^\mu (0) | \hat{\pi}_a (p_1) \rangle = - i \epsilon_{abc} \mathrm{Tr}(\sigma_c \bm{X}^\prime_V )  (p_1 + p_2)^\mu\,,
\end{equation}
where $\epsilon_{abc}$ is the $SU(2)$ structure constant. This result can be derived by writing the expression of the vector current at leading order in ChPT. A straightforward calculation then yields for the decay rate
\begin{align}
 &\; \Gamma ( \pid_a \to \pid_b\,  f \bar{f} ) \simeq \;  \frac{m_{\pid_a}^5}{192 \pi^3} \left[ \epsilon_{abc} {\rm Tr}(\sigma_c \bm{X}'_V)\right]^2  \int_{4y^2}^{(1-x)^2} dt \left( 1 - \frac{4y^2}{t} \right)^{1/2}  \\ 
\times &\; \left[ \lambda(1, x^2, t)^{3/2} \left( 1 + \frac{2y^2}{t} \right) \left( |C_{VV}^f|^2 + |C_{VA}^f|^2 \right) + 6 \lambda(1, x^2, t)^{1/2}  y^2 ( 2 + 2x^2 - t) |C_{VA}^f|^2 \right]\,, \nonumber
\end{align}
where $x = m_{\pid_b} / m_{\pid_a}$, $y = m_f/m_{\pid_a}$, $t = q^2 / m_{\pid_a}^2$ where $q^2 = (p_f + p_{\bar{f}})^2 = (p_1 - p_2)^2$. Recall that $\lambda$ was defined in Eq.~\eqref{eq:defs_Kallen}. Here $f$ was assumed to be a charged lepton or neutrino. If the fermion mass can be neglected, the integral $\int_0^{(1 - x)^2} dt\, \lambda (1, x^2, t)^{3/2}$ equals $\simeq (16/5) (1-x)^5$ for $x \to 1$, implying that the decay width is suppressed by $(m_{\hat{\pi}_a} - m_{\hat{\pi}_b})^5$, whereas it equals $\simeq 1/4$ for $x \to 0$. The pion mass splittings are expected to be small for $N = 2$, because they do not arise at leading order in the quark masses. Sizeable splittings are more likely for larger $N$, in which case one simply replaces $\sigma_a/2$ by the corresponding generator $T_a$ and $\epsilon_{abc} \to f_{abc}\,$.

\subsection{Dark pion mixing}\label{app:dark_pion_mix}

Here we discuss, in the model of Eq.~\eqref{eq:Lagr}, the mixings of the angular ($\tilde{a}$) and radial ($\phi$) modes of $\Phi$ with the dark pions, which generate splittings among the dark pion masses. Writing $\Phi = (v_\Phi + \phi/\sqrt{2})e^{i \tilde{a}/ (\sqrt{2}\hspace{0.1mm}v_\Phi)}$, we can express the dressed quark mass matrix as
\begin{equation}
\widetilde{\bm{m}}_\psi = \tilde{A}^\dagger \Big( \bm{m}_\psi + \frac{\bm{\zeta} \phi}{\sqrt{2}}   \Big) \tilde{A}
\end{equation}
with $\tilde{A} \equiv \mathrm{diag}\,(e^{-i\tilde{a}/(2\sqrt{2}\hspace{0.1mm}v_\Phi)}, \,e^{+i\tilde{a}/(2\sqrt{2}\hspace{0.1mm}v_\Phi)})$. In the basis where the (undressed) quark mass matrix is diagonal, this becomes $\widetilde{\bm{m}}_{\psi^\prime} = U_L^\dagger \tilde{A}^\dagger \Big( \bm{m}_\psi + \frac{\bm{\zeta} \phi}{\sqrt{2}}   \Big) \tilde{A}\, U_R$. In this basis, the quark charges under $U(1)'$ are $\bm{X}^\prime_L$ and $\bm{X}^\prime_R$. This is the basis the dark ChPT Lagrangian corresponds to, so we have
\begin{equation}
\mathcal{L}_{2} = \frac{f_{\hat{\pi}}^2}{4}\mathrm{Tr} [(D_\mu \Sigma)^\dagger D^\mu \Sigma] + \frac{\hat{B}_0 f_{\hat{\pi}}^2}{2} \mathrm{Tr} (\widetilde{\bm{m}}_{\psi^\prime}\Sigma^\dagger + \mathrm{h.c.} ) 
\end{equation}
where $\Sigma = e^{i \hat{\Pi} / f_{\hat{\pi}}}$, $\hat{\Pi} = \hat{\pi}^a \sigma_a$, and $D_\mu \Sigma = \partial_\mu \Sigma + i g_D \hat{Z}^\prime_\mu \bm{X}^\prime_L \Sigma - i g_D \hat{Z}^\prime_\mu \Sigma \bm{X}^\prime_R$. The constant $\hat{B}_0$ was defined in Eq.~\eqref{eq:B0hat}. Upon going to the vacuum, $\langle \Sigma \rangle = \bm{1}_2$, we learn that the mass of the $\hat{Z}^\prime$ is
\begin{equation}
\hat{M}^2_{Z^\prime} = 2 g_D^2 \big[ x_\Phi^2 v_\Phi^2 +  \mathrm{Tr}( \bm{X}^{\prime \dagger}_A \bm{X}^\prime_A ) f_{\hat{\pi}}^2 \big],
\end{equation}
where the first contribution comes from the $\Phi$ VEV and the second one from the dark QCD vacuum. Furthermore, there is mass mixing between $\tilde{a}$, $\phi$, and the dark pions. 

To see these effects in a more transparent way, let us focus on the simple ($CP$-conserving) case $y_2 = m_2 = 0$. Then we find
\begin{equation}
\mathrm{Tr}(\bm{X}^{\prime \dagger}_A \bm{X}^\prime_A ) f_{\hat{\pi}}^2  =  (x_1 - x_2)^2  \frac{ y_1^2 v_\Phi^2 }{2 (y_1^2 v_\Phi^2 + m_1^2)}  f_{\hat{\pi}}^2\,,
\end{equation}
while by expanding the mass term in $\mathcal{L}_2$ we obtain the mass Lagrangian
\begin{equation}\label{eq:mass_atilde}
- \frac{1}{2} \hat{B}_0 \sqrt{y_1^2 v_\Phi^2 + m_1^2} \begin{pmatrix} \tilde{a} & \hat{\pi}_1 & \hat{\pi}_3 \end{pmatrix} \begin{pmatrix} \frac{y_1^2 f_{\hat{\pi}}^2}{2(y_1^2 v_\Phi^2 + m_1^2)} &  \frac{ -\, y_1 m_1 f_{\hat{\pi}}}{\sqrt{2}(y_1^2 v_\Phi^2 + m_1^2)} &  \frac{ - \,y_1^2 v_\Phi f_{\hat{\pi}}}{\sqrt{2}(y_1^2 v_\Phi^2 + m_1^2) } \\ & 1 & 0 \\ & &  1  \end{pmatrix} \begin{pmatrix} \tilde{a} \\ \hat{\pi}_1 \\ \hat{\pi}_3 \end{pmatrix}\,.
\end{equation}
This mass matrix has zero determinant, signaling the presence of a null eigenvalue, which corresponds to the physical mode eaten by the $Z'$. Notice that $\tilde{a}$ mixes with both $\hat{\pi}_1$ and $\hat{\pi}_3$, as in general allowed by $CP$ conservation. The two nonzero eigenvalues are $\hat{B}_0 \sqrt{y_1^2 v_\Phi^2 + m_1^2}$ and $\hat{B}_0 \sqrt{y_1^2 v_\Phi^2 + m_1^2}\, [ y_1^2 (v_\Phi^2 + f_{\hat{\pi}}^2/2) + m_1^2 ] / ( v_\Phi^2 y_1^2 + m_1^2)$. The same result can be obtained by considering the four-dark quark operators generated by $Z'$ exchange in Eq.~\eqref{eq:Leff},
\begin{equation}
\mathcal{L}_{\rm eff} \simeq - \frac{g_D^2}{2M_{Z^\prime}^2} (J_{D})^2 \quad \to \quad -\frac{f_{\hat{\pi}}^2}{4(x_1 - x_2)^2 v_\Phi^2} \Big[\sum_b \mathrm{Tr}(\sigma_b \bm{X}^\prime_A) \partial_\mu \hat{\pi}_b \Big]^2 \,.
\end{equation}
This means that the kinetic matrix of $CP$-odd pions is corrected as
\begin{equation}
\frac{1}{2} \begin{pmatrix} \partial \hat{\pi}_1 & \partial \hat{\pi}_3 \end{pmatrix} \begin{pmatrix} 1 - C_1^2 &\; - C_1 C_3 \\ &\; 1 - C_3^2 \end{pmatrix} \begin{pmatrix} \partial \hat{\pi}_1 \\ \partial \hat{\pi}_3 \end{pmatrix} \,, \qquad C_{1,3} \equiv \frac{f_{\hat{\pi}} \mathrm{Tr}(\sigma_{1,3} \bm{X}^\prime_A)}{\sqrt{2}(x_1 - x_2) v_\Phi}\,,
\end{equation}
which after diagonalization by a rotation leaves $1$ and $1 - C_1^2 - C_3^2$ on the diagonal. Thus, by redefining the wavefunction of one of the pions we obtain the physical squared masses for $y_2 = m_2 = 0\,$,
\begin{equation}
\hat{B}_0 \sqrt{y_1^2 v_\Phi^2 + m_1^2} \,,\qquad \hat{B}_0 \sqrt{y_1^2 v_\Phi^2 + m_1^2}\;  (1 + C_1^2 + C_3^2) = \hat{B}_0 \sqrt{y_1^2 v_\Phi^2 + m_1^2}  \left( 1 + \frac{y_1^2 f_{\hat{\pi}}^2/2}{y_1^2 v_\Phi^2 + m_1^2} \right)\,,
\end{equation}
in agreement with the ChPT picture. However, one also needs to remember that in ChPT the pion masses receive corrections from NLO operators, such as
\begin{equation}
\mathcal{L}_4 \supset \frac{c_7 \hat{B}_0^2 y_1^2}{(4\pi)^2} \Big( \mathrm{Tr}[\widetilde{\bm{m}}_{\psi^\prime} \Sigma^\dagger - \mathrm{h.c.} ] \Big)^2 \to - \frac{1}{2} \frac{c_7 \hat{B}_0^2 y_1^2}{2\pi^2} \begin{pmatrix} \tilde{a} & \hat{\pi}_1 & \hat{\pi}_3 \end{pmatrix} \begin{pmatrix} \frac{y_1^2 v_\Phi^2}{2(y_1^2 v_\Phi^2 + m_1^2)} & 0 & - \frac{v_\Phi}{\sqrt{2}f_{\hat{\pi}}} \\  & 0 & 0 \\ &  & \frac{y_1^2 v_\Phi^2 + m_1^2}{y_1^2 f_{\hat{\pi}}^2} \end{pmatrix} \begin{pmatrix} \tilde{a} \\ \hat{\pi}_1 \\ \hat{\pi}_3 \end{pmatrix}\,,
\end{equation}
where the last expression applies to the $y_2 = m_2 = 0$ case. In general, this mass matrix should be added to the one in Eq.~\eqref{eq:mass_atilde} and leads to an $\sim m_{\hat{\pi}}^2/(4\pi f_{\hat{\pi}})^2$ correction to the mass of $\hat{\pi}_3$. 

Finally, there is mixing between $\phi$ and $\hat{\pi}_2$. In ChPT this reads
\begin{equation}
- \frac{1}{2} \begin{pmatrix} \phi & \hat{\pi}_2 \end{pmatrix} \begin{pmatrix} m_\phi^2 & \quad - \frac{\hat{B}_0 f_{\hat{\pi}}}{\sqrt{2}}  \frac{y_1 m_1}{\sqrt{m_1^2 + y_1^2 v_\Phi^2}} \vspace{1mm} \\ & \quad \hat{B}_0 \sqrt{m_1^2 + v_1^2 v_\Phi^2} \end{pmatrix} \begin{pmatrix} \phi \\ \hat{\pi}_2 \end{pmatrix}\,,
\end{equation}
leading in the limit $m_\phi^2 \gg m_{\hat{\pi}}^2$ to a negative correction to the squared mass of $\hat{\pi}_2$ of the form $\delta m^2_{\hat{\pi}_2} \simeq - \tfrac{\hat{B}_0^2 y_1^2 f_{\hat{\pi}}^2}{2m_\phi^2} \frac{ m_1^2}{m_1^2 + y_1^2 v_\Phi^2}\,$. The same result is obtained by considering the four-dark quark operators mediated by $\phi$ exchange,
\begin{equation}
\mathcal{L}_{\rm eff} \supset \frac{1}{4m_\phi^2} \left( \overline{\psi}^{\,\prime}_L \bm{\zeta}^\prime \psi^\prime_R + \overline{\psi}^{\,\prime}_R \bm{\zeta}^{\prime\,\dagger} \psi^\prime_L \right)^2 \quad \to \quad \frac{\hat{B}_0^2 f_{\hat{\pi}}^2}{16 m_\phi^2} \Big[ \sum_b \mathrm{Tr} [ i \sigma_b (\bm{\zeta}^\prime - \bm{\zeta}^{\prime\,\dagger})  ] \hat{\pi}_b \Big]^2\,,
\end{equation}
which after specializing to the $y_2 = m_2 = 0$ scenario and using the first relation in the second line of Eq.~\eqref{eq:m2_0} indeed yields the same $\delta m^2_{\hat{\pi}_2}$.

\subsection{Dark meson production in FCNC decays of SM mesons}\label{app:FCNC}
Starting from the four-fermion Lagrangian in Eq.~\eqref{eq:FCNC_4fermion} the amplitude for the process $\text{Had}_1\to \text{Had}_2 + \hat{\pi}_b\,$, where $\mathrm{Had}_{1,2}$ are SM hadrons, can be obtained from factorization as
\begin{equation}\label{eq:factorized}
\langle \hat{\pi}_b\, \text{Had}_2  | \mathcal{H}_{\rm eff}^{\rm FCNC} | \text{Had}_1 \rangle = \frac{i \, g^2}{64\pi^2}  V_{tj}^\ast V_{ti} \mathcal{K}_t \frac{p_{\hat{\pi}}^\mu}{f_a^{(b)}} \,\langle\, \text{Had}_2 | \bar{d}_j \gamma_\mu P_L d_i |\text{Had}_1  \rangle~,
\end{equation}
where the hadronic matrix elements can be written in terms of momentum-dependent form factors. We focus on $B$ meson decays and make use of available results obtained from light-cone QCD sum rules~\cite{Ball:2004ye,Gubernari:2018wyi}. For decays involving dark pions we find
\begin{align}
\Gamma(B \to K \hat{\pi}_b) =&\; \frac{m_B^3}{64\pi} f_0^K (m_{\hat{\pi}}^2)^2 \left| \frac{g^2 V_{t s}^\ast V_{tb}}{64\pi^2 f_a^{(b)}} \mathcal{K}_t \right|^2 \Big(1 - \frac{m_K^2}{m_B^2} \Big)^2 \lambda_{B K \hat{\pi}}^{1/2}\,, \\
\frac{\Gamma(B \to K^\ast \hat{\pi}_b)}{\Gamma(B \to K \hat{\pi}_b)} =&\; \frac{A_0^{K^\ast} (m_{\hat{\pi}}^2)^2}{f_0^K(m_{\hat{\pi}}^2)^2} \frac{\lambda_{B K^\ast \hat{\pi}}^{3/2}}{\Big( 1 - \frac{m_K^2}{m_B^2} \Big)^2 \lambda_{B K \hat{\pi}}^{1/2}}\,,
\end{align}
where $f_0^K \equiv f_0^{B\to K}$, and so forth. For decays involving dark vector mesons, $\text{Had}_1\to \text{Had}_2 + \hat{\rho}_b$, we write similarly to Eq.~\eqref{eq:factorized}
\begin{equation}\label{eq:factorized_2}
\langle \hat{\rho}_p\, \text{Had}_2  | \mathcal{H}_{\rm eff}^{\rm FCNC} | \text{Had}_1 \rangle = \frac{i \,g^2}{128\pi^2} V_{tj}^\ast V_{ti} \mathcal{K}_t\, \hat{g}_Z \varepsilon_{Zp}\, \epsilon_\mu^\ast (p_{\hat{\rho}}) \,\langle\, \text{Had}_2 | \bar{d}_j \gamma^\mu P_L d_i |\text{Had}_1  \rangle~,
\end{equation}
leading to
\begin{equation}
\Gamma(B \to K \hat{\rho}_p) = \frac{m_B^3}{64\pi} f_{+}^K (m_{\hat{\rho}}^2)^2 \left| \frac{g^2 V_{t s}^\ast V_{tb}}{128\pi^2} \mathcal{K}_t \frac{\hat{g}_Z \varepsilon_{Zp}} { m_{\hat{\rho}} } \right|^2 \lambda_{B K \hat{\rho}}^{3/2}\,,
\end{equation}
and
\begin{align}
&\,\Gamma(B \to K^\ast \hat{\rho}_p) =  \frac{m_B^3}{64\pi} \left| \frac{g^2 V_{t s}^\ast V_{tb}}{128\pi^2} \mathcal{K}_t \frac{\hat{g}_Z \varepsilon_{Zp}} { m_{\hat{\rho}} } \right|^2  \nonumber \\ 
\times&\, \frac{\lambda_{B K^\ast \hat{\rho}}^{3/2}}{4}\bigg[8 V^{K^\ast}\hspace{-0.5mm} (m_{\hat{\rho}}^2)^2 \frac{m_{\hat{\rho}}^2}{(m_B + m_{K^\ast})^2}  + A_1^{K^\ast}\hspace{-0.5mm} (m_{\hat{\rho}}^2)^2 \frac{(m_B + m_{K^\ast})^2}{m_{K^\ast}^2}\bigg( 1 + \frac{12m_{K^\ast}^2 m_{\hat{\rho}}^2}{m_B^4 \lambda_{BK^\ast \hat{\rho}}} \bigg) \label{eq:B0_Kstar_rhohat_width} \\ 
&\,\qquad\qquad + A_2^{K^\ast}\hspace{-0.5mm} (m_{\hat{\rho}}^2)^2 \frac{m_B^4 \lambda_{B K^\ast \hat{\rho}}}{m_{K^\ast}^2(m_B + m_{K^\ast})^2} - 2 A_1^{K^\ast}\hspace{-0.5mm} (m_{\hat{\rho}}^2) A_2^{K^\ast}\hspace{-0.5mm} (m_{\hat{\rho}}^2) \frac{ m_B^2 - m_{K^\ast}^2 - m_{\hat{\rho}}^2 }{m_{K^\ast}^2 } \bigg] \nonumber\,.
\end{align}
Notice that for light $\hat{\rho}$ the contribution of $V^{K^\ast}$ is suppressed by $m_{\hat{\rho}}^2 m_{K^\ast}^2/m_B^4 \ll 1$ relative to $A_{1,2}^{K^\ast}$. We approximate the form factors with their values at zero momentum transfer, taken from the light-cone QCD sum rule analysis of Ref.~\cite{Gubernari:2018wyi},
\begin{equation}
f_{+}^K = f_{0}^K = 0.27\,, \quad A_{0}^{K^\ast} = 0.31\,,\;\; V^{K^\ast} = 0.33\,,\;\; A_{1}^{K^{\ast}} = 0.26\,,\;\; A_{2}^{K^{\ast}} = 0.24\,.
\end{equation}
We are now ready to evaluate the expected rates, which are provided in Eq.~\eqref{eq:B_K_pihat} for decays involving dark pions, and Eqs.~\eqref{eq:B0_K_rhohat} and~\eqref{eq:B0_Kstar_rhohat} for decays involving dark vector mesons. 

Finally, for very light dark pions ($m_{\hat{\pi}} < m_K - m_\pi$) kaon decays are kinematically open, for instance
\begin{align}
\Gamma(K^+ \to \pi^+ \hat{\pi}_b) =&\; \frac{m_K^3}{64\pi} f_0^{K\to\, \pi} (m_{\hat{\pi}}^2)^2 \left| \frac{g^2 V_{t d}^\ast V_{ts}}{64\pi^2 f_a^{(b)}} \mathcal{K}_t \right|^2 \Big(1 - \frac{m_\pi^2}{m_K^2} \Big)^2 \lambda_{K \pi \hat{\pi}}^{1/2}\,,
\end{align}
where $f_0^{K\to\, \pi} (0) = 0.97$ from a lattice QCD determination~\cite{Carrasco:2016kpy}, leading to
\begin{equation}
\mathrm{BR}(K^{+} \to \pi^+ \hat{\pi}_b  ) \approx 3.9 \times 10^{-11}\, \bigg( \frac{1\;\mathrm{PeV}}{f_a^{(b)}} \bigg)^2 \bigg( \frac{\mathcal{K}_t}{10} \bigg)^2  \lambda_{K \pi \hat{\pi}}^{1/2} \,.
\end{equation}

\section{Dark baryon dark matter}\label{app:DM}
Assuming the simplest scenario with $N = 2$ dark flavors and $N_d = 3$ dark colors, the lightest baryons are the dark proton $\hat{p} = (\psi_{1}^\prime \psi_{1}^\prime \psi_{2}^\prime)$ and dark neutron $\hat{n} = (\psi_{1}^\prime \psi_{2}^\prime \psi_{2}^\prime)$. Taking $m_{\psi_1^\prime} < m_{\psi_2^\prime}$ for definiteness, $\hat{p}$ is the lightest baryon and therefore stable, whereas the dark neutron can decay rapidly as $\hat{n} \to \hat{p} f \bar{f}$ (with $f$ a SM fermion). The $\hat{p}$ is a candidate for asymmetric dark matter (DM), but it is well known that direct detection (DD) bounds are very strong if DM interacts with the SM via the $Z$ portal. To evaluate the corresponding cross section we start from the piece of Eq.~\eqref{eq:4f_SM_dark} that reads $\mathcal{L}_{\rm eff} \supset -\, \overline{\psi}^{\,\prime} \gamma^\mu X^\prime_V \psi^\prime \sum_f C_{VV}^f \bar{f} \gamma_\mu f$, from which we obtain the interaction of the dark proton with the SM valence quarks $u,d$,
\begin{equation}
 - (X^\prime_V)_{\hat{p}}\; \bar{\hat{p}} \gamma^\mu \hat{p} \sum_{f \, = \, u, d} C_{VV}^f \bar{f} \gamma_\mu f\,.
\end{equation}
Here we defined $(X^\prime_V)_{\hat{p}} \equiv 2 (X^\prime_V)_{\psi_1^\prime} + (X^\prime_V)_{\psi_2^\prime}$, which we take to be an $O(1)$ number. Notice that both the mass and kinetic mixings contribute to $C_{VV}^f$ and therefore to DD, as expected. The cross section for spin-independent scattering of $\hat{p}$ on SM nucleons $N$ is
\begin{equation}
\sigma_{\rm SI}^{\hat{p} N} = \frac{1}{\pi} \left(   \frac{ m_N m_{\hat{p}}}{m_{\hat{p}} + m_N} \right)^2 \frac{(X^\prime_V)_{\hat{p}}^2}{A^2} \left[ Z \sum_{f = u,d} F_{V_f}^p C_{VV}^f + (A - Z) \sum_f F_{V_f}^n C_{VV}^f \right]^2\,,
\end{equation}
where $F_{V_u}^p = F_{V_d}^n = 2$ and $F_{V_d}^p = F_{V_u}^n = 1$. Setting the kinetic mixing to zero, we find
\begin{align}
\sigma_{\rm SI}^{\hat{p} N} \simeq &\; \frac{m_N^2}{16\pi} \left(   \frac{ m_{\hat{p}}}{m_{\hat{p}} + m_N} \right)^2 (X^\prime_V)_{\hat{p}}^2 \left( \frac{ 2Z (1 - 2 \hat{s}_W^2) - A}{A} \right)^2 \left( \frac{\delta \hat{M}^2}{\hat{M}_Z^2} \right)^2 \frac{\hat{g}_Z^2 g_D^2}{M_{Z^\prime}^4} \\
\approx&\; 5.6 \times 10^{-43}\;\mathrm{cm}^2 \left(   \frac{ m_{\hat{p}}}{m_{\hat{p}} + m_N} \right)^2 (X^\prime_V)_{\hat{p}}^2  \left( \frac{\delta \hat{M}^2 / \hat{M}_Z^2}{10^{-2}} \right)^2 \left( \frac{g_D}{0.25} \right)^2 \left( \frac{60\;\mathrm{GeV}}{M_{Z^\prime}} \right)^4 \,,\nonumber
\end{align}
where in the second line we have assumed scattering on Xenon. A comparison with experimental results~\cite{Billard:2021uyg} indicates that, for $m_{\hat{p}} \lesssim 3\;\mathrm{GeV}$, a spin-independent cross section $\lesssim 5 \times 10^{-42}\;\mathrm{cm}^2$ is still allowed. Therefore, in this range of $\hat{p}$ mass (roughly corresponding to $f_{\hat{\pi}} \lesssim 0.3\;\mathrm{GeV}$) a $Z^\prime$ satisfying $M_{Z^\prime} < M_Z$ could be a viable mediator between asymmetric dark baryon DM and the SM. 

Another aspect requiring attention is that this type of DM has sizeable self-interactions $\hat{p}\hat{p}\to \hat{p}\hat{p}$, for which $\sigma / m$ cannot be too large due to astrophysical bounds from the Bullet cluster and smaller scales, see e.g.~Ref.~\cite{Tulin:2017ara} for a review. Attempting to calculate the scattering cross section, which sensitively depends on the details of the model, goes beyond the scope of this paper. See for instance Ref.~\cite{Cline:2013zca} for previous work repurposing results of lattice calculations performed in SM QCD.

\bibliographystyle{JHEP}
\bibliography{darkshower}

\end{document}